\documentclass{aa}
\usepackage{graphicx}
\usepackage{txfonts}
\usepackage{natbib}
\usepackage{mathrsfs}
\usepackage{longtable}
\usepackage[shortlabels]{enumitem}
\usepackage{siunitx}
\usepackage{epstopdf}
\usepackage{hyperref}	
\hypersetup{colorlinks=true,linkcolor=blue,citecolor=blue,filecolor=blue,urlcolor=blue}
\usepackage{comment}
\usepackage{txfonts}

\begin{document}
\title{Galactic Archaeology with asteroseismic ages. II.\\ Confirmation of a delayed gas infall using Bayesian\\
analysis based on MCMC methods}
\author {E. Spitoni\inst{1} \thanks {email to: spitoni@phys.au.dk}
  \and K. Verma\inst{1}  \and
  V. Silva  Aguirre\inst{1}  \and F. Calura
  \inst{2}}
\institute{Stellar Astrophysics Centre, Department of Physics and
  Astronomy, Aarhus University, Ny Munkegade 120, DK-8000 Aarhus C,
  Denmark \and I.N.A.F. - Osservatorio Astronomico di Bologna, Via Gobetti 93/3, 40129 Bologna, Italy}

\date{Received xxxx / Accepted xxxx}

\abstract
{  With the wealth of information from large surveys and observational campaigns in the contemporary era, 
it is critical to properly exploit the data to constrain the parameters of Galactic chemical evolution 
models and quantify the associated uncertainties.}
{ We aim at constraining the two-infall chemical evolution models for the solar annulus using the measured 
chemical abundance ratios and seismically inferred age of stars in the APOKASC sample.
In the revised two-infall chemical evolution models by \citet{spitoni2019}, a significant delay of 
$\sim 4.3$ Gyr has been invoked between the two episodes of gas accretion. In this work, we wish to  test its robustness and statistically confirm/quantify the delay.}
{ For the first time, a Bayesian framework based on Markov Chain Monte Carlo methods has been used for fitting the 
two-infall chemical evolution models to the data.}
{ In addition to fitting the data for stars in the APOKASC sample, our best fit models also reproduce  other important observational constraints of the chemical evolution of the disk: i) present 
day stellar surface density; ii) present-day supernova and star formation rates; iii) the metallicity 
distribution function; and iv) solar abundance values.  We found a significant delay between the two gas 
accretion episodes for various models explored with different values for the star formation efficiencies. 
The values for the delay lie in the range $4.5-5.5$ Gyr.}
{ The results suggest that the APOKASC sample carries the signature of delayed gas-rich merger, with
the dilution as main process determining the shape of low-$\alpha$ stars in the abundance ratios space.}


\titlerunning{Delayed gas infall and Bayesian analysis}

\authorrunning{Spitoni et al.}

\maketitle

\section{Introduction}

The purpose of Galactic Archaeology is to unveil the formation and evolution of our Galaxy by interpreting 
signatures imprinted in the observed chemical abundances and kinematics of resolved stellar populations. 
This is typically done through the proper exploitation of the observational stellar data to constrain models 
of Galactic chemical evolution. The contemporary wealth of data from big surveys and observational 
campaigns, e.g. spectroscopic properties from the Apache Point Observatory Galactic Evolution Experiment
project \citep[APOGEE;][]{Majewski:2017ip}, kinematic properties from the `fossil' record of old stellar
populations as provided by the Gaia mission \citep[DR2;][]{gaia2_2018}, and precise seismic ages from the
{\it Kepler} satellite \citep{borucki2009}, offer an unprecedented opportunity to test models of galaxy 
formation and evolution.

The analysis of the APOGEE data \citep{Nidever:2014fj, hayden2015} suggested the existence of a clear distinction between two sequences of disc stars in the [$\alpha$/Fe] versus [Fe/H] abundance ratio space: the so-called high-$\alpha$  and low-$\alpha$ sequences.
 This dichotomy has been also confirmed by  the Gaia-ESO  survey \citep[e.g.,][]{RecioBlanco:2014dd,RojasArriagada:2016eq,RojasArriagada:2017ka} and the AMBRE project \citep{Mikolaitis:2017gd}.

In several theoretical models of the Galactic discs evolution it has been proposed that this  bimodality is strictly connected to a delayed gas  accretion episode of primordial composition. 
For instance, a late second accretion phase after a prolonged period with a quenched star formation rate (SFR) has been 
suggested by the dynamical models presented by \citet{noguchi2018}. Moreover, the AURIGA simulations presented 
by \citet{grand2018} clearly point out that a bimodal distribution in the [Fe/H]-[$\alpha$/Fe] plane is a 
consequence of a significantly lowered gas accretion rate at ages between 6 and 9 Gyr.
In the framework of cosmological hydrodynamic simulations of Milky Way like galaxies, \citet{buck2020} stated that a 
bimodal $\alpha$-sequence is a generic consequence of a gas-rich merger at some time in galaxy's evolution. As 
also suggested by \citet{spitoni2019}, the merger gives rise to the low-$\alpha$ sequence by 
bringing pristine metal-poor gas in the system which dilutes the metallicity of interstellar medium while keeping 
[$\alpha$/Fe] abundance almost unchanged (as first proposed in a cosmological model by \citealt{calura2009}).

The model presented by \citet{spitoni2019} (hereafter ES19) also includes precise stellar ages provided by asteroseismology to constrain the 
  chemical evolution  of the solar neighbourhood.
   The ES19 model is an updated version of the classical "two-infall" of \citet{chiappini1997}, in which 
 an early fast gas accretion episode gives rise to the high-$\alpha$ sequence, and at a 
later Galactic time, the low-$\alpha$ sequence is created by a different infall event characterized by a 
longer time-scale of accretion.
The predictions of the revised ``two-infall" models were compared with the measured 
chemical abundance ratios  \citep{pinso2014} and seismically inferred age of stars in the APOKASC catalogue 
\citep[APOGEE + {\it Kepler} Asteroseismology Science Consortium;][]{victor2018}.  ES19 model was capable 
of reproducing the APOKASC data assuming a disc component dissection based on chemistry \citep[see][]{victor2018}, 
i.e. the sample was divided in two distinct groups called `high-$\alpha$' and `low-$\alpha$' sequences.
The most important result of ES19 was that a significant delay of $\sim$ 4.3 Gyr 
between the two infall episodes was required to reproduce the measured stellar abundances and seismically inferred 
ages.

In ES19  the choice of  free parameters, i.e. the two infall time scales, the corresponding star formation 
efficiencies and the delay between the two infall episodes of the model was made to qualitatively reproduce the observed [$\alpha$/Fe] versus [Fe/H] abundance ratios.
In this article we  present a  quantitative study of the free parameters using a Bayesian analysis.
Probabilistic data analysis has transformed scientific research in the past decade. In particular, Bayesian 
analysis based on Markov Chain Monte Carlo (MCMC) methods have been used in several different areas of astrophysics
including cosmology (\citealt{dunkley2005}), cosmic rays (\citealt{putze2010}), active galactic nuclei 
(\citealt{reynolds2012}), Milky Way dwarf satellites (\citealt{ural2015}), semi-analytical models of galaxy 
formation (\citealt{kampa2008,henri2009,henri2013}), and stellar nucleosynthesis (\citealt{cescutti2018}), among
other. Recently, MCMC methods have been used in testing Galactic chemical evolution models 
\citep[see e.g.][]{cote2017,rybi2017,philcox2018,frankel2018,belfiore2019}.

In this paper, we present the first attempt 
to perform a detailed study of the key parameters 
which regulate the evolution of the solar neighbourhood by means of a  match between a Bayesian MCMC method and the  two-infall chemical evolution model. 
The goal is to test the findings of ES19 by quantitatively inferring the delay 
between the two accretion episodes without imposing any stellar data separation based on chemical abundances.

The paper is organised as follows: in Section \ref{s:apokasc}   the observational data used in the Bayesian analysis  is presented,  in Section \ref{s:chemmod} we briefly recall the main characteristics of two 
infall model, in Section \ref{fitting} we describe the fitting method and also perform a preliminary test, in 
Section \ref{results_astero} we present our results, and finally in Section \ref{conc} we summarize our conclusions.

\section{The APOKASC sample}\label{s:apokasc}

 In this work we use a Bayesian framework based on MCMC 
methods to fit state-of-the-art models of Galactic chemical evolution to the observed 
chemical abundance ratios and asteroseismic ages of stars in the  updated APOKASC (APOGEE+ {\it Kepler} Asteroseismology Science Consortium) sample presented by \citet{victor2018}.

The sample is  composed by 1197 red giants spanning out to 2 kpc in the solar annulus with stellar properties determined combining the photometric, spectroscopic, and asteroseismic observables in the BAyesian STellar Algorithm \citep[{\tt BASTA},][]{silvaaguirre2015,silvaAguirre2017} framework.
The sample is also characterized by  precise kinematic information available  from the first DR of Gaia \citep{linde2016,gaia2016} and The Fourth US Naval Observatory CCD Astrograph  (UCAC-4) catalogue \citep{zach2013}.
Here, it is assumed that $\alpha$ abundances are given by the sum of the individual Mg and Si abundances \citep{salaris2018}.

As in ES19, in  the present work  we do not consider  the so-called ``young $\alpha$ rich'' (Y$\alpha$R) stars.
The origin of these stars is still uncertain  and  two different scenarios have been proposed:  either  they  are  objects  migrated  from the Galactic bar \citet{chiappini2015}  or
  evolved blue stragglers \citep{martig2015,chiappini2015,yong2016,jofre2016}.

\section{The revised two-infall  model by ES19 }\label{s:chemmod}
In this Section we recall the main assumptions  and characteristics of the  revised two-infall chemical evolution model  proposed by
ES19. 
A few details on the model are provided, including the parametrization of the most basic physical processes (e. g. infall and star formation), as well as the stellar nucleosynthesis prescriptions used in the work. 
 
\subsection{The chemical evolution model prescriptions}
In ES19   the authors revised   the classical
``two-infall'' chemical evolution model in order to  reproduce  the
data from updated APOKASC catalogue by \citet{victor2018} which were  chemically dissected in
high-$\alpha$ and  low-$\alpha$ stellar sequences.
 From the  precise stellar ages determined via asteroseismology,
a clear age
difference   emerged in the solar annulus between  high-$\alpha$ and  low-$\alpha$ stars.
 The low-$\alpha$ sequence age distribution peaks at $\sim$ 2 Gyr,
 whereas the high-$\alpha$  one does so at $\sim$ 11 Gyr.

In ES19 the Milky Way disc is assumed to be formed by two distinct accretion
episodes of gas. The gas infall rate is expressed by the following expression,
\begin{equation}
\mathcal{I}_i(t,i)=(\mathcal{X}_i)_{inf} \left[ \mathcal{N}_1 \, e^{-t/ \tau_{1}}+ \theta(t-t_{{\rm max}}) \mathcal{N}_2 \, e^{-(t-t_{{\rm max}})/ \tau_{2}} \right],
\label{a}
\end{equation}
where $\tau_{1}$  is the  time-scale for the formation of the
high-$\alpha$ sequence which was fixed at a value of 0.1 Gyr and 
$\tau_{2}$ is the  time-scale  for the formation of the low-$\alpha$
disc phase which was fixed at a value of 8 Gyr.
We remind the reader that the $\theta$ in the equation above is the Heaviside step function. 
$(\mathcal{X}_i)_{inf}$ is the abundance by mass of the element $i$ in the
infalling gas which is assumed to have primordial composition, whereas $t_{{\rm max}}$=4.3
Gyr is the time of the maximum infall rate on the  second accretion episode, i.e. it indicates the delay of 
the beginning of the second infall. ES19 emphasised the importance of the $t_{{\rm max}}$  value in order to properly reproduce the APOKASC data. Finally, the coefficients $\mathcal{N}_1 $ and $\mathcal{N}_2 $ are obtained by imposing a
 fit to the observed current total surface mass density in the solar
 neighbourhood adopting the  relations

\begin{equation}
\mathcal{N}_1 =\frac{\sigma_1}{\tau_{1} \left(1- e^{-t_G/\tau_{1}}\right)},
\label{S1}
\end{equation}

\begin{equation}
\mathcal{N}_2 =\frac{\sigma_2}{\tau_{2} \left(1-
 e^{-(t_G-t_{{\rm max}})/\tau_{2}} \right)},
\label{S2}
\end{equation}
where $\sigma_1$ and $\sigma_2$ are the present
day total surface mass density of the high-$\alpha$ and low-$\alpha$ sequence stars,
respectively; $t_G$ is the Age of the Milky Way.

The star formation rate is expressed as the \citet{kenni1998} law,
\begin{equation}
\psi(t) =\nu \sigma_{g}^{k}(t),
\label{k1}
\end{equation}
where $\nu$ is the star formation efficiency (SFE),  $\sigma_g$ is the surface
gas density, and $k = 1.5$ is the exponent. 
In ES19  the adopted SFE is constant during the whole Galactic life and fixed at the value of $\nu=1.3$ Gyr$^{-1}$. 
However, different infall episodes could in principle be characterized by different SFEs.
In fact, in the two-infall model by
\citet{grisoni2017,grisoni2018} the SFEs associated to the high and
low $\alpha$ sequences are  different: $\nu_1=2$ Gyr$^{-1}$ and
$\nu_2=1$ Gyr$^{-1}$.

 We adopt the \citet{scalo1986}  initial stellar mass function (IMF), constant in
time and space.

\subsection{Nucleosynthesis prescriptions and solar values}
As for the nucleosynthesis prescriptions for  Fe, Mg and Si, ES19 adopted
the ones suggested by \citet{francois2004}. For a detailed description we
refer to ES19.
This set of yields  has been widely used in the literature
\citep{cescutti2007,  spitoni2011, Mott2013, spitoni2015,
  spitoni2017, spitoni2D2018,vincenzo2019} and turned out to be able to reproduce the main features of the solar neighbourhood.

We have adopted the photospheric values of \citet{asplund2005} as our solar reference abundances to be consistent with the APOGEE data release \citep{gperez2016}.

\section{The fitting method}\label{fitting}

 Bayesian analysis based on MCMC methods 
has transformed scientific research in the past decade.
Since 
there are already several text books and reviews on Bayesian statistics 
\citep[see e.g.][]{jaynes2003,gelman2013} and MCMC methods 
\citep[see e.g.][]{brooks2011,sharma2017,hogg2018,speagle2019}, we only describe 
them briefly and highlight the aspects specific to the problem at hand. 

In the context of parameter estimation, Bayes' theorem provides a way to update
the parameters based on any new  available data. In other words, it enables the calculation 
of the posterior probability distribution of the parameters given the new data,
\begin{equation}
 P({\bf \Theta}|{\bf x})=\frac{ P({\bf \Theta})}{ P({\bf x})}P({\bf x}|{\bf \Theta}),
 \label{eq:bayes}
\end{equation}

where ${\bf x}$ represents the set of observables, ${\bf \Theta}$ the set of model parameters,
$P({\bf x}|{\bf \Theta}) \equiv \mathscr{L}$ the likelihood (i.e. probability of observing the data given the 
model parameters), $P({\bf \Theta})$ the prior (i.e. probability of the model parameters 
before seeing the data, ${\bf x}$), and $P({\bf x})$ represents the evidence (i.e. total 
probability of observing the data). The evidence is a normalizing constant and can be calculated 
by integrating the likelihood over all model parameters. In the current study, 
${\bf x} = \{[\alpha/{\rm Fe}], \ [{\rm Fe}/{\rm H}], \ {\rm age}\}$ and 
${\bf \Theta} = \{\tau_1, \ \tau_2, \ t_{\rm max},  \ \sigma_2 / \sigma_1\}$ 
are examples of the set of observables and model parameters, respectively.

\begin{figure}
\begin{centering}
\includegraphics[scale=0.45]{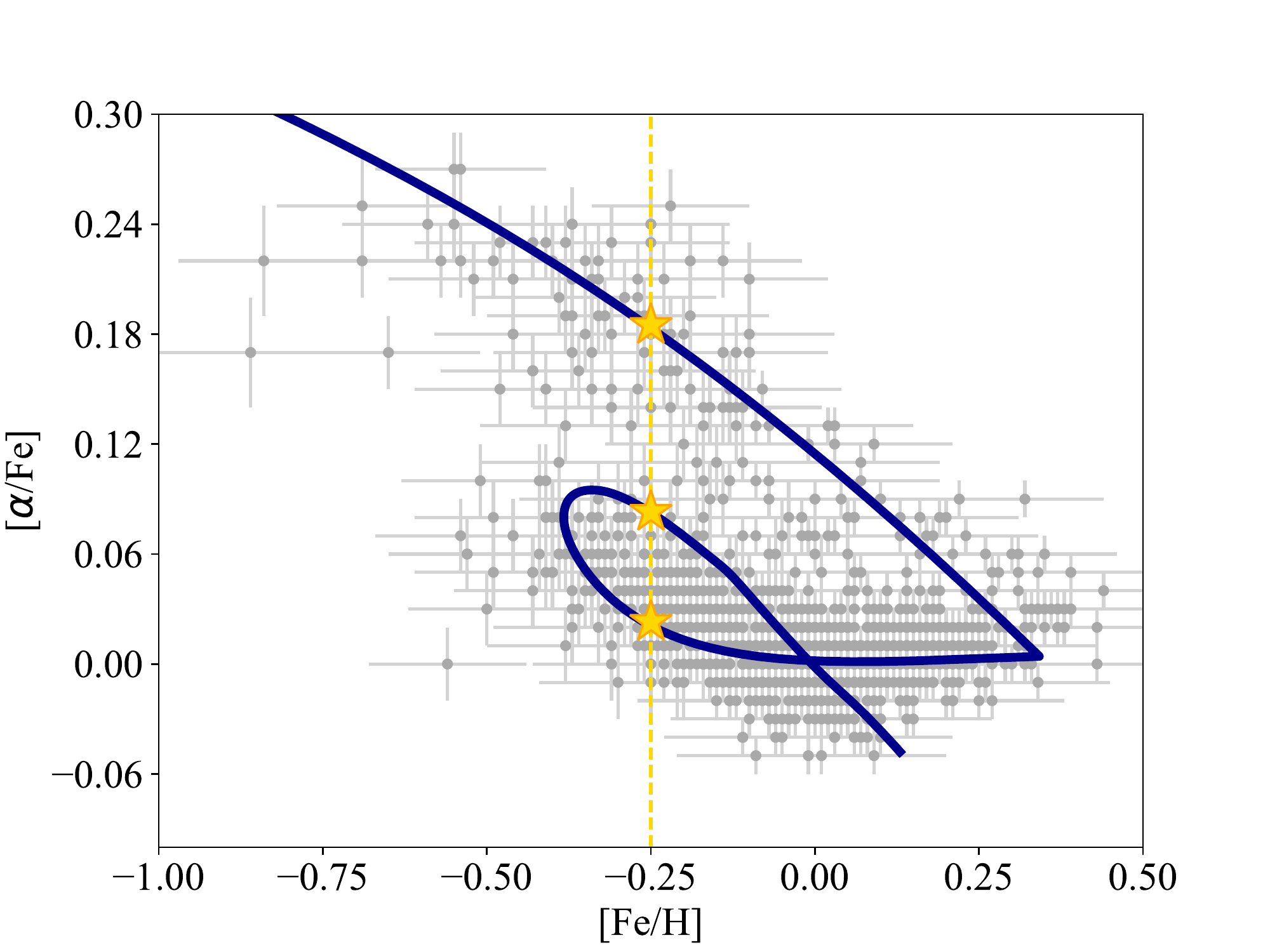}
\caption{Observed [$\alpha$/Fe] versus [Fe/H] abundance ratios of the APOKASC sample  (grey points)  presented by \citet{victor2018}, compared with the two-infall chemical evolution model characterized by $\tau_1= 0.1$ Gyr,   $\tau_2= 8$ Gyr, $\sigma_2=64$ M$_{\odot}$ pc$^{-2}$,   $\sigma_2/\sigma_1$=4, $\nu=1.3$ Gyr$^{-1}$. The three yellow points   clearly show  that the predicted chemical evolution curve is multi-valued in this plane (they exhibit three different [$\alpha$/Fe] values for the same [Fe/H] abundance ratio). }
\label{ES19}
\end{centering}
\end{figure}

To define the likelihood in eq.~(\ref{eq:bayes}) we assume that the uncertainties on the 
observables are normally distributed. In that case, the logarithm of the likelihood can be
written as,
\begin{equation}
\ln  \mathscr{L}  =-\sum_{n=1}^N\ln\left(\left(2 \pi \right)^{d/2} \prod_{j=1}^d \sigma_{n,j}\right)
-\frac{1}{2}\sum_{n=1}^N\sum_{j=1}^d\left(\frac{x_{n,j}-\mu_{n,j}}{\sigma_{n,j}} \right)^2,
\label{eq:likelihood}
\end{equation}
where $N$ is the number of stars in the sample and $d$ is the number of observables available.
The quantities $x_{n,j}$ and $\sigma_{n,j}$ are respectively the measured value of $j^{\rm th}$ 
observable and its uncertainty for $n^{\rm th}$ star. In principle, the quantity $\mu_{n,j}$ 
is the model value of $j^{\rm th}$ observable for $n^{\rm th}$ star, in practice however it is
tricky to define (as explained below).

To define $\mu_{n,j}$, we first consider the case of fitting the data only in the [Fe/H]-$[\alpha/{\rm Fe}]$ 
plane. As can be seen in Fig. \ref{ES19}, the curve predicted by the two-infall model in this plane is 
multivalued, i.e. there are more than one values of $[\alpha/{\rm Fe}]$ for certain values of [Fe/H]. As a result, 
it becomes ambiguous to associate an observed data point in the
[Fe/H]-$[\alpha/{\rm Fe}]$ plane to a point on the curve, making it difficult to define $\mu_{n,j}$. 
 In 
this study, we associate a data point to the closest point on the curve. Given a data point $x_{n,j}$, this is done by defining the following  function,

\begin{equation}
S_n \equiv \min_i\left\{\sqrt{\sum_{j=1}^{2} \left( \frac{x_{n,j}-\mu_{n,j,i}}{\sigma_{n,j}} \right)^2}\right\}= \sqrt{\sum_{j=1}^{2} \left( \frac{x_{n,j}-\mu_{n,j,i'}}{\sigma_{n,j}} \right)^2},
\label{eq:distance1}
\end{equation}
where $i$ runs over the different points on the curve. Hence, the closest point on the curve  is $\mu_{n,j} = \mu_{n,j,i'}$. This definition can be easily generalized 
to define $\mu_{n,j}$ for an arbitrary number of observables by modifying Eq.~\ref{eq:distance1},

\begin{equation}
S_n \equiv \min_i\left\{\sqrt{\sum_{j=1}^{d} \left( \frac{x_{n,j}-\mu_{n,j,i}}{\sigma_{n,j}} \right)^2}\right\}=\sqrt{\sum_{j=1}^{d} \left( \frac{x_{n,j}-\mu_{n,j,i'}}{\sigma_{n,j}} \right)^2}.
\label{eq:distance2}
\end{equation}

The computation of the posterior also requires the specification of priors on the model parameters (see 
eq.~\ref{eq:bayes}). Here, we discuss the priors on ${\bf \Theta} = \{\tau_1, \ \tau_2, \ t_{\rm max}, 
\ \sigma_2 / \sigma_1\}$  and justify their choices in the current study.
\begin{itemize}
    \item First infall time-scale  $\tau_1$: in the classical two-infall model  \citep{chiappini1997} the first gas infall 
    is characterized by a short time-scale of accretion and it has been fixed at the value of $\tau_1$= 1 Gyr. More recently, in order to reproduce the AMBRE thick disc, 
    \citet{grisoni2017} suggested a smaller value:  $\tau_1=0.1$ Gyr. 
    In the current study, we set a uniform prior
    on $\tau_1$, exploring the range $0 < \tau_{1} < 7  \mbox{ Gyr}$.

    \item The second infall time-scale, $\tau_2$,  is 
    connected to a slower accretion episode. We set a uniform prior on  $\tau_2$ 
    exploring the range $0 < \tau_2 < 28  \mbox{ Gyr}$, since there is no reason for $\tau_2$ being limited to the age of the Universe.
     \item The delay $t_{\rm max}$: we set a uniform prior 
    exploring the range $0 < t_{\rm max} < 14  \mbox{ Gyr}$, which extends all the way to the age 
    of the Universe.

    \item Present-day two surface mass density ratio, $\sigma_2/\sigma_1$: there are still large discrepancies 
    in the estimates of the thick disc surface density quoted in the literature, which contribute to large 
    uncertainties in the estimates of $\sigma_2/\sigma_1$. For instance,  
    \citet[][and references therein]{nesti2013} claimed that the ratio between low- and high-$\alpha$ 
    sequence stars should be  around 10. On the other hand, \citet{fuhr2017} derived a local mass density 
    ratio between thin and thick disc stars of 5.26, which becomes as low as 1.73 after correction for the 
    difference in the scale height. While studying APOGEE stars \citet{mac2017} found that the relative 
    contribution of low- to high-$\alpha$ is 5.5. Bearing in mind these uncertainties, we set a uniform prior 
    on the mass density ratio, exploring the range $1 < \sigma_2/\sigma_1 < 50$ (assuming that the low-$\alpha$ 
    component is more massive than the high-$\alpha$ one).
\end{itemize}

Finally, we sampled the posterior probability distribution defined by eq.~(\ref{eq:bayes}) using an affine 
invariant MCMC ensemble sampler \citep{goodman2010,foreman}. This was accomplished using the publicly 
available code "\textit{emcee}: the mcmc hammer"
\footnote{\href{https://emcee.readthedocs.io/en/stable/}{https://emcee.readthedocs.io/en/stable/};
\href{https://github.com/dfm/emcee}{https://github.com/dfm/emcee}}. We initialized the chains with 100 walkers 
and ran the sampler for 1000 steps (see below for the details).

\subsection{Testing the method}\label{section_only_abu}

\begin{figure}
\begin{centering}
\includegraphics[scale=0.55]{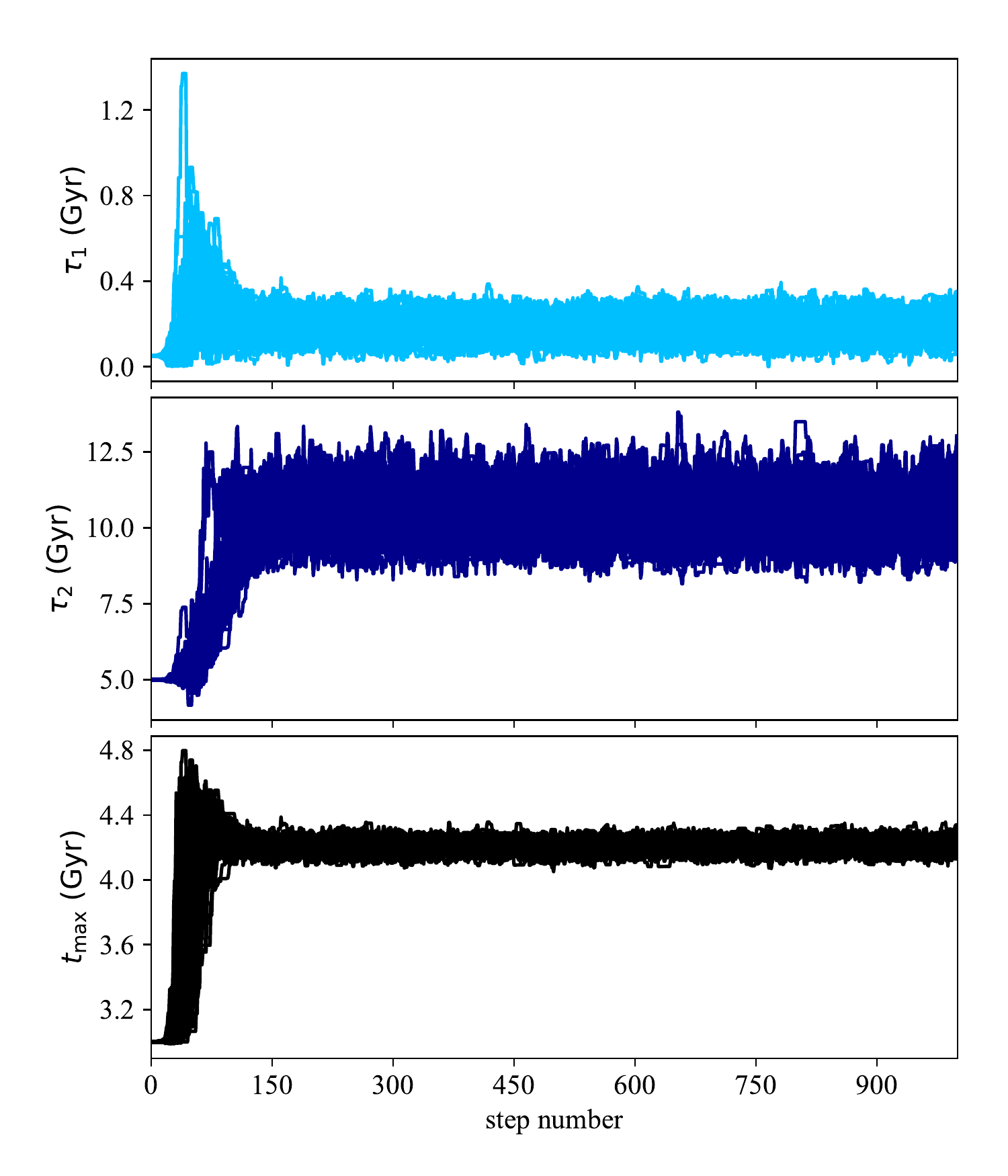}
\caption{ Convergence of the three free model parameters composed by the set  ${\bf \Theta} = \{\tau_1, \ \tau_2, \ t_{\rm max}\}$  (see model details in Section \ref{section_only_abu})   as  a function of number of steps.}
\label{conv1}
\end{centering}
\end{figure}

\begin{figure}
\begin{centering}
\includegraphics[scale=0.47]{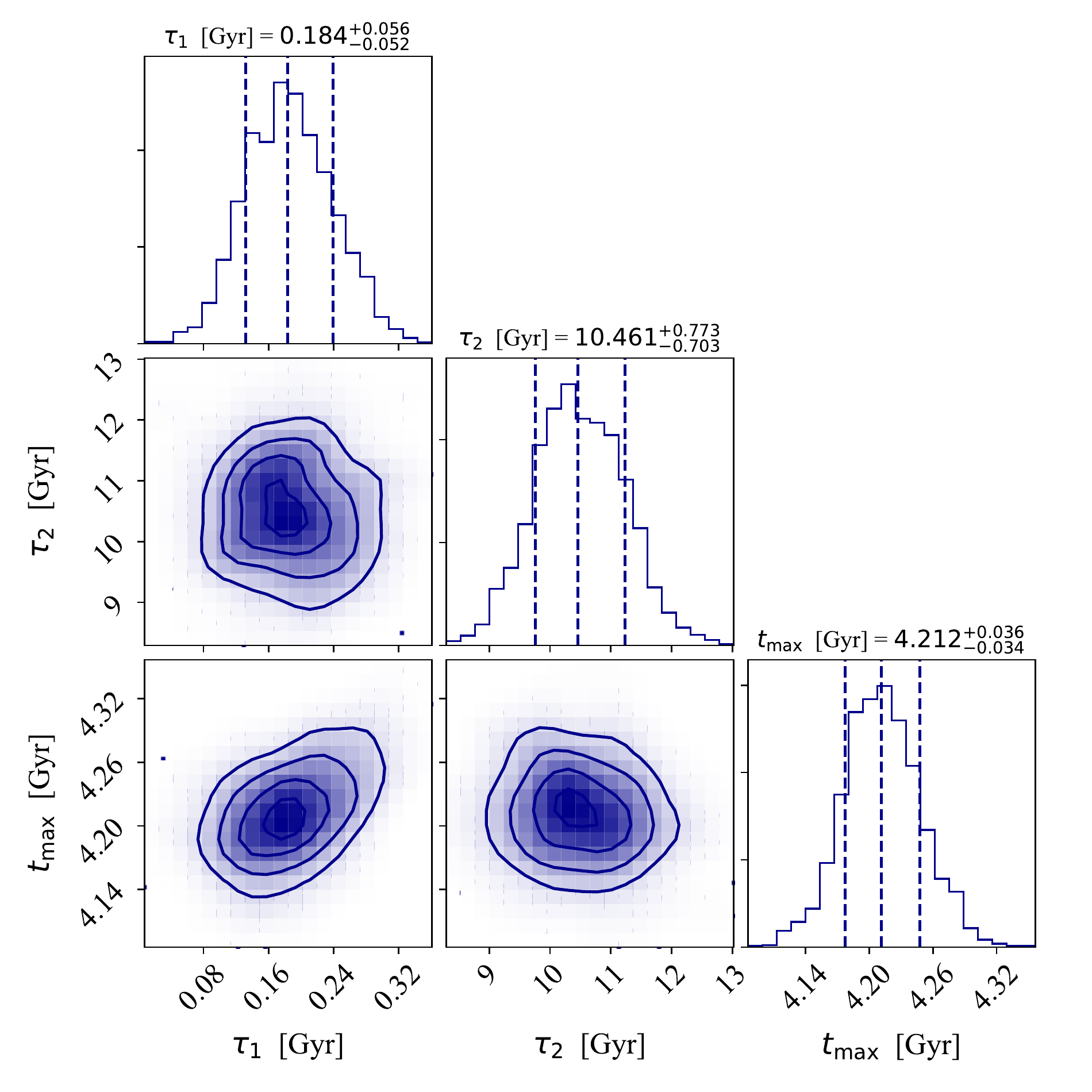}
\caption{  Corner plot showing the posterior PDFs for the same model set as in Fig. \ref{conv1} obtained by fitting [$\alpha$/Fe] and  [Fe/H]
  abundances from the \citet{victor2018} sample. For each parameter, the median, 16$^{\rm th}$  and 84$^{\rm th}$ percentiles of
  the posterior PDF are plotted above the 
  marginalised PDF. }
\label{corner1}
\end{centering}
\end{figure}

\begin{figure}
\begin{centering}
\includegraphics[scale=0.45]{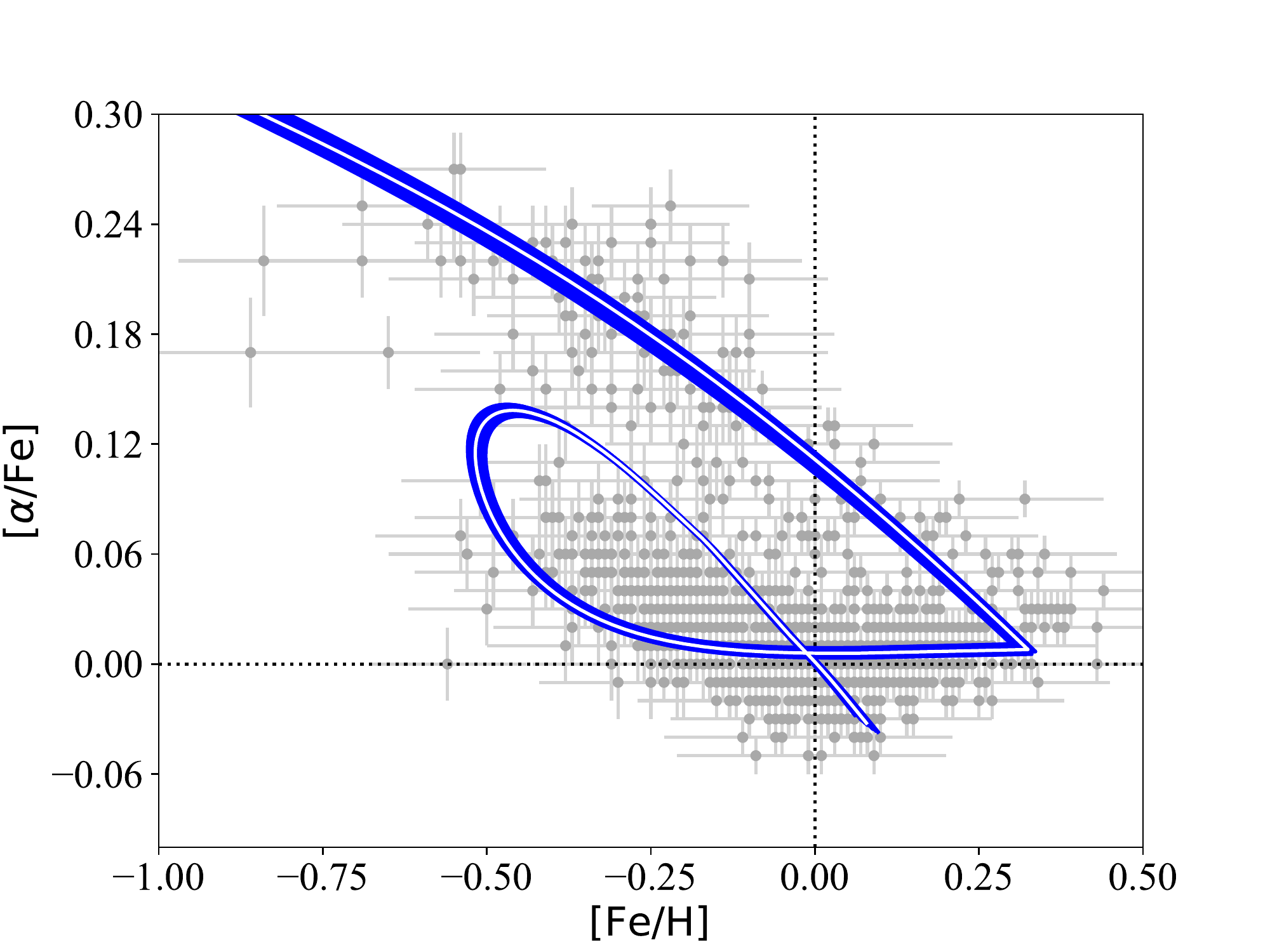}
\caption{Observed [$\alpha$/Fe] versus [Fe/H] abundance ratios
  presented by \citet{victor2018}  (grey points), compared with  model results constrained only by abundance ratio data  with three free
  parameters, composed by the set  ${\bf \Theta} = \{\tau_1, \ \tau_2, \ t_{\rm max}\}$  (see in Section \ref{section_only_abu}). 
  The blue lines correspond to  one hundred walkers at the last  step, and the thin white line indicates the best fit model. }
\label{alpha1}
\end{centering}
\end{figure}

In this Section we test the method  showing results   when  model free parameters are constrained just by  chemical abundance ratios, i.e.  the   likelihood calculation is based only on
[$\alpha$/Fe] and [Fe/H] abundance ratio data.
We recall that in ES19, the presence  of a significant delay between the two infall episodes ($\sim$ 4.3 Gyr)  was a crucial assumption to properly reproduce the APOKASC data.

We consider three free  model  parameters:  the infall
time-scales of accretion  $\tau_{1}$ (first gas infall),  $\tau_{2}$
(second infall),  and  the delay
 $t_{{\rm max}}$ between the start of the two infall episodes. The SFE has been fixed to the value of  $\nu= 1.3$ Gyr$^{-1}$, whereas the  present-day surface gas density of the high- and
 low-$\alpha$ sequences are  8 M$\odot$ pc$^{-2}$ and  64 M$\odot$ pc$^{-2}$, respectively  as suggested by the ES19 best model adopting \citet{nesti2013} prescriptions.

It should be noted that the number of steps considered in a MCMC
calculation can have a significant impact on the results
\citep{goodman2010,foreman}.
In Fig. \ref{conv1} we show the evolution of 100 walkers as a function of the number of steps. As it can be seen,   the chains have converged already after 200 steps, thus ensuring the robustness of the results.
The posterior   probability density function (PDF)  of the model parameters is presented in Fig.  \ref{corner1}. The best fit model parameters
are:   $\tau_1 =0.184^{+0.056}_{-0.052}$ Gyr,
$\tau_2=10.461^{+0.773}_{-0.703}$ Gyr and $t_{{\rm max}} = 4.212^{+0.036}_{-0.034}$ Gyr.
In Fig. \ref{alpha1} we show  the abundance ratios [$\alpha$/Fe] versus [Fe/H]  predicted by 100 walkers computed at the last  time-step
of the MCMC steps (at the 1000$^{\rm th}$ step). In this plot  the thickness of the model curve represents the uncertainty.

In Fig. \ref{alpha1} we notice that the best model is similar to the
one of ES19 (see their Fig. 2),  
and the associated "loop" feature
in the [$\alpha$/Fe] and [Fe/H] space  related to the low-$\alpha$ sequence is retained.
The newly constrained  free parameter values by the MCMC algorithm are similar to 
the ones of ES19,  and  we obtain an almost identical value for the delay $t_{{\rm max}}$ with a difference of $\Delta t_{{\rm max}}=0.088$ Gyr.

\begin{figure}
\begin{centering}
\includegraphics[scale=0.37]{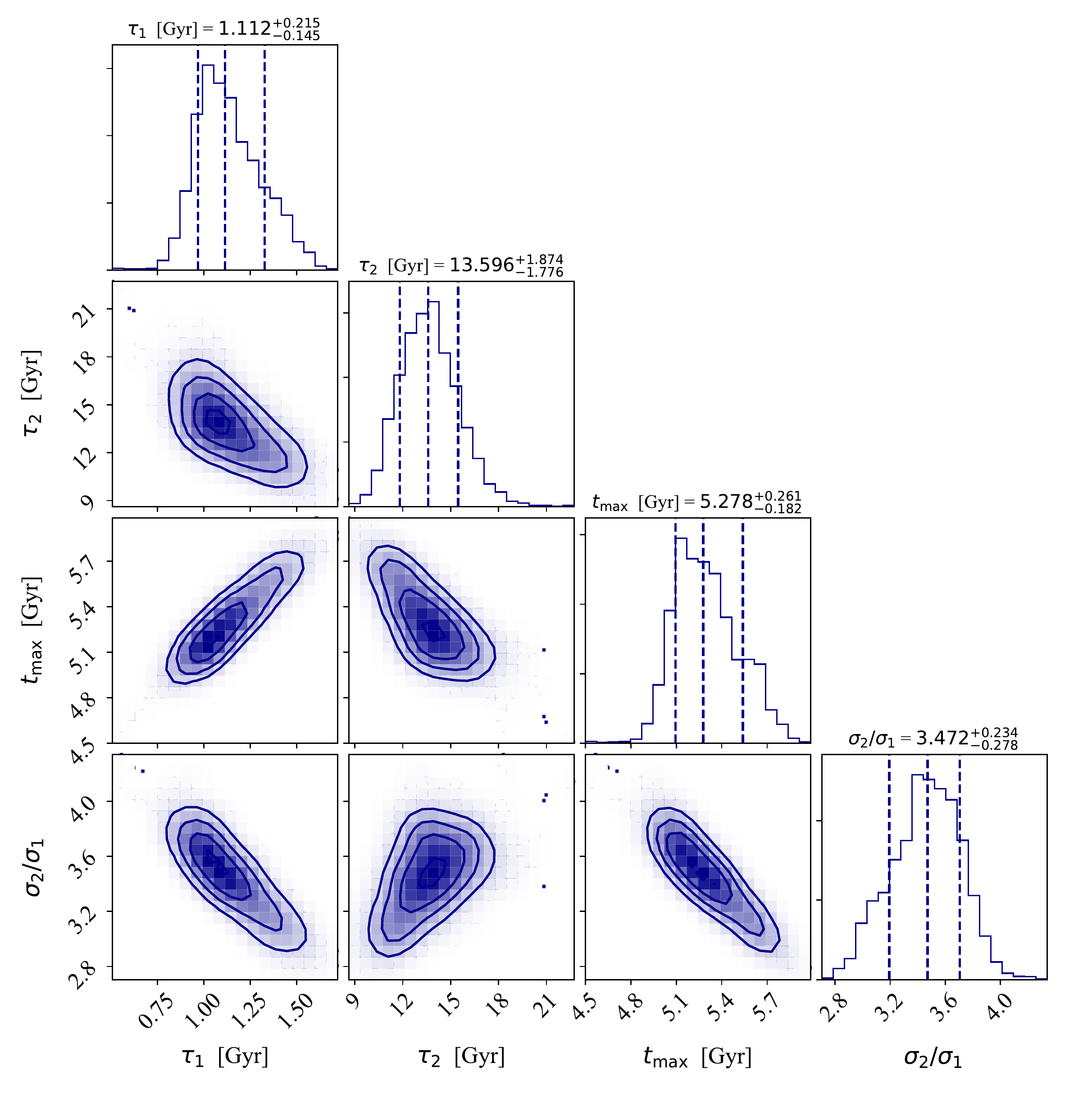}
\caption{Corner plot showing the posterior PDFs of  model M1 in which the SFE has been fixed at the value of $\nu=1.3$ Gyr$^{-1}$. In this case the best fit of the four dimensional parameter space ${\bf \Theta} = \{\tau_1, \ \tau_2, \ t_{\rm max}, \ \sigma_2 / \sigma_1\}$  is  obtained by fitting [$\alpha$/Fe],  [Fe/H] and ages
  of the APOKASC sample (see model details in Section \ref{results_astero}). The median, 16$^{\rm th}$  and 84$^{\rm th}$ percentiles of
  the posterior PDF are plotted 
  for each parameter above the marginalised PDF. }
\label{4par_corner_M1}
\end{centering}
\end{figure}

\begin{figure}
\begin{centering}
\includegraphics[scale=0.43]{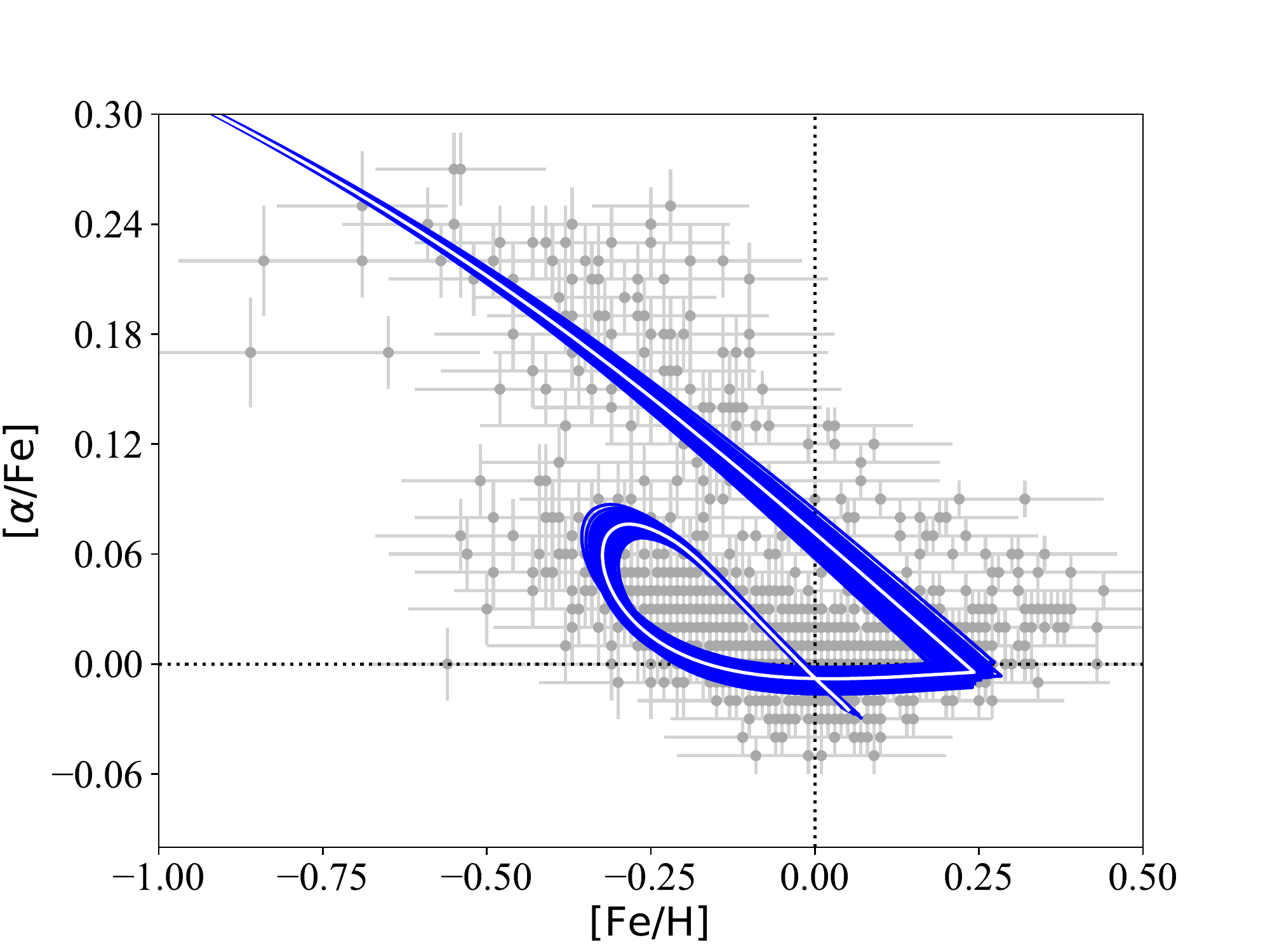}
\includegraphics[scale=0.43]{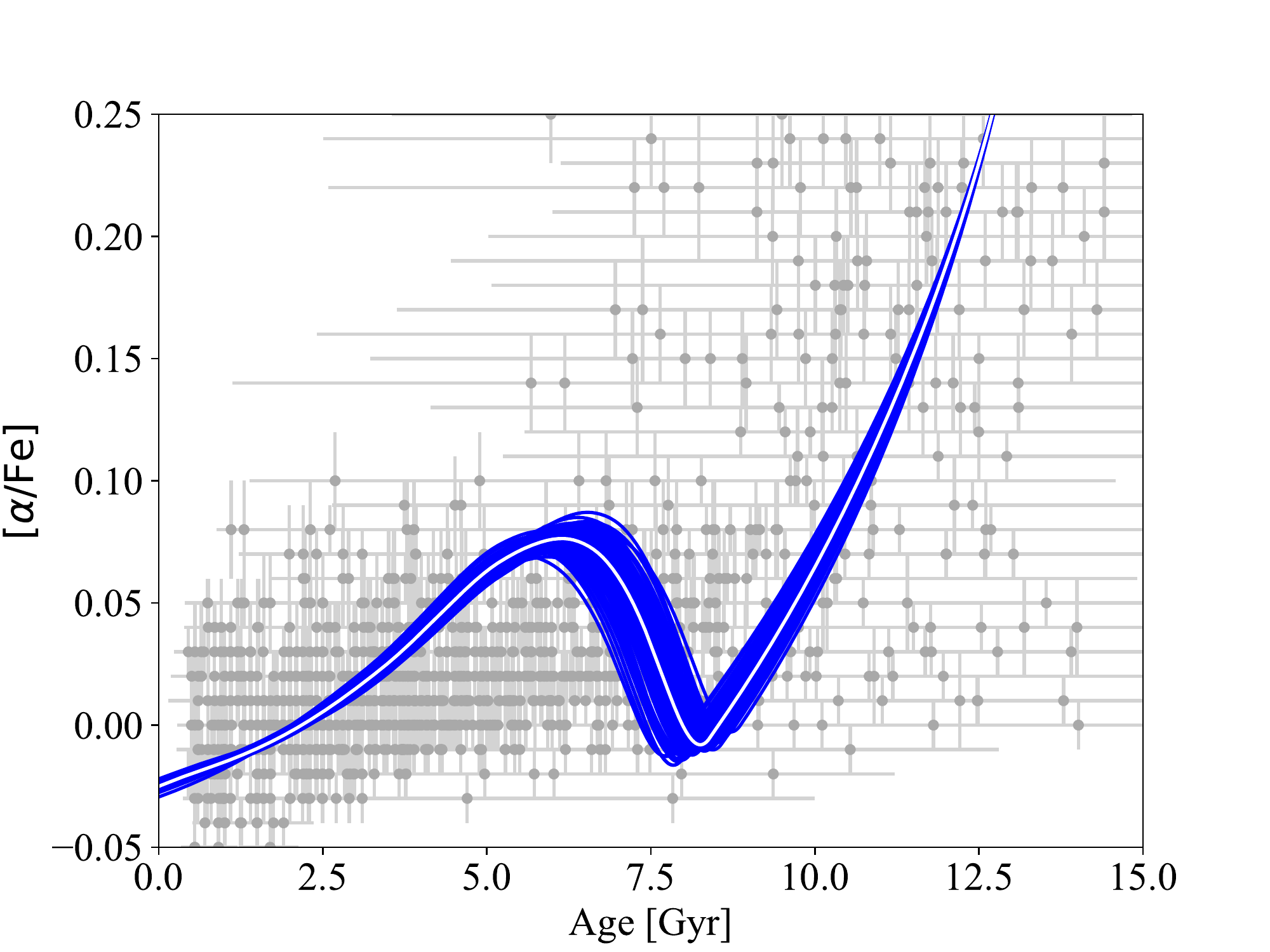}
\includegraphics[scale=0.43]{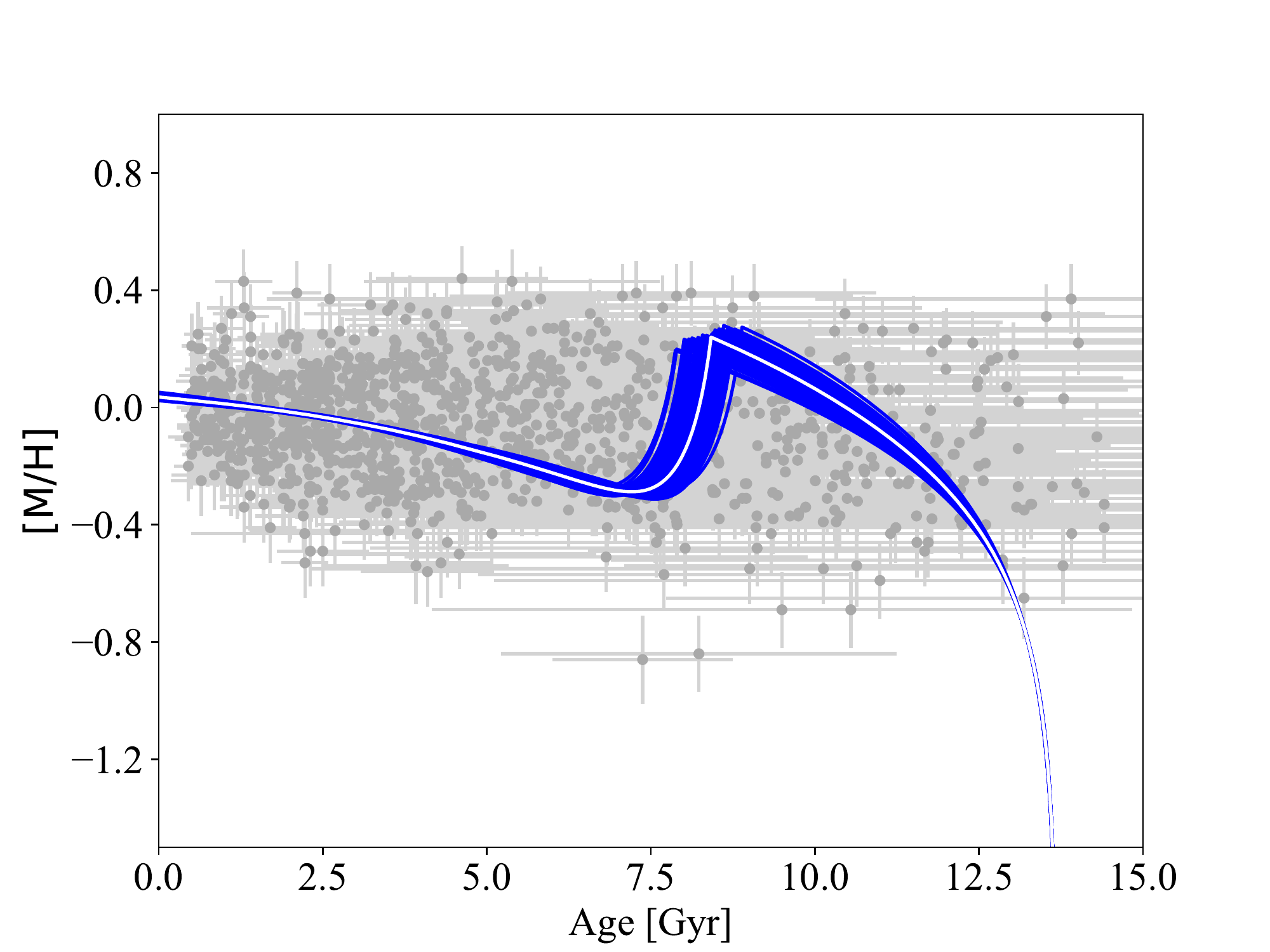}
\caption{{\it Upper panel}: Same as Fig. \ref{alpha1}  except that the model M1 is also constrained by means of the APOKASC stellar ages. The best fit model parameters of the set  ${\bf \Theta} = \{\tau_1, \ \tau_2, \ t_{\rm max},  \ \sigma_2 / \sigma_1\}$ was  computed by means of the PDFs of Fig. \ref{4par_corner_M1}. In the {\it Middle} and {\it Lower panel} the temporal evolution of  [$\alpha$/Fe] and the age-metallicity relation are reported, respectively,  for the same models and data of the upper panel. }
\label{alpha_fe_4param_M1}
\end{centering}
\end{figure}
\begin{figure*}
\begin{centering}
\includegraphics[scale=0.5]{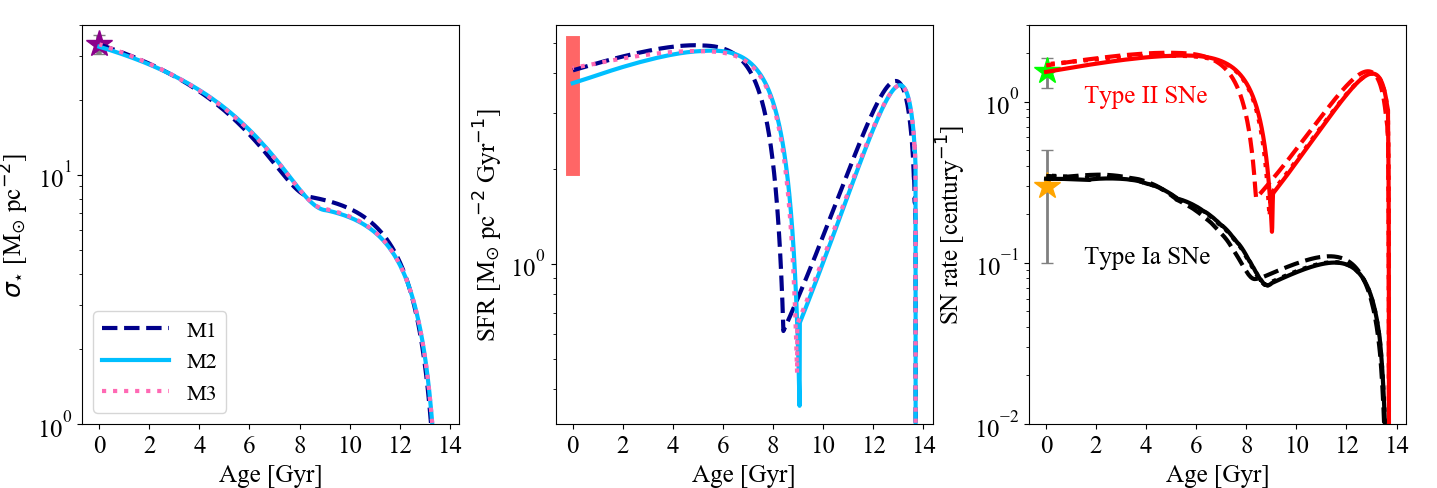}
\caption{ {\it Left panel}: Stellar surface mass density $\sigma_{\star}$ evolution predicted by the best fit models M1 (long dashed line), M2 (solid line) and M3 (dotted line).  The purple star indicates the observed present-day value given by \citet{mckee2015}.
{\it Middle panel}: same as the left panel for SFR time evolution. The red shaded area  indicates the measured range in the solar annulus suggested by \citet{prantzos2018}. {\it Right panel}: evolution of the Type Ia SN (black lines) and Type II SN (red lines) rates predicted by models M1, M2 and M3 for the whole Galactic disc. The yellow  star stands for the  observed Type Ia SN rate observed by \citet{cappellaro1997}, whereas the green one stands for Type II SN rates observed by \citet{li2010}. }
\label{SFR_SN}
\end{centering}
\end{figure*}

\begin{table}[htp]
\begin{center}
\caption{Observed solar chemical abundances compared with the theoretical ones predicted by the best  models M1, M2 and M3 (see text for model details) constrained by  chemical abundance ratios and stellar ages of the APOKASC sample.}
\label{tab1}
\begin{tabular}{c|cccc}

  \hline
  \hline

 Abundance &{\it Observations} &  \multicolumn{3}{c}{\it Models}\\
 $\log$($X$/H)+12 & \citet{asplund2005}&M1&M2&M3\\
 
 & [dex]&[dex]&[dex] &[dex] \\
\hline
\\
Fe &7.45$\pm$0.05&7.31& 7.31 & 7.33\\

Si &7.51$\pm$0.04& 7.41&7.40 & 7.42\\

Mg& 7.53$\pm$0.09 &7.44&  7.43&7.45\\

 \hline
\end{tabular}
\end{center}
\end{table}

\section{Results} \label{results_astero}

In this Section we show model predictions  in presence of the new dimension provided by
asteroseismology, i.e. precise stellar ages. The  free parameters are determined by fitting [$\alpha$/Fe], [Fe/H] and stellar ages provided by the APOKASC sample.
We note that, in this analysis,  we do not assume the disc component dissection between high-$\alpha$ and low-$\alpha$ stellar
sequences by \citet{victor2018} based on the chemistry. 

It should be noted that the total surface mass density is another important local key observable. \citet{mckee2015} suggested that the total surface density including the thin and thick components in the solar neighborhood  should be 47.1 $\pm$ 3.4 M$_{\odot} \mbox{ pc}^{-2}$ and that the total local surface density of stars is 33.4 $\pm$ 3 M$_{\odot} \mbox{ pc}^{-2}$. In this study, we  use the value of total surface density (sum of high-$\alpha$ and low-$\alpha$) of 47.1 $\pm$ 3.4 M$_{\odot} \mbox{ pc}^{-2}$ as provided by \citet{mckee2015} because they also quote constraint on the stellar mass content.

In contrast to \citet{spitoni2019}, we do not impose the present-day total low-$\alpha$ sequence surface density (see eq. \ref{S2}) but  instead use the total surface mass densities ($\sigma_{tot} =  \sigma_1+\sigma_2$) to be constant as given by the \citet{mckee2015} study.
Recalling that $\sigma_2/\sigma_1$ is the ratio between the low-$\alpha$  and high-$\alpha$  present-day total surface mass density, we have,
\begin{equation} 
 \sigma_{tot}=  \sigma_2 \left( 1+\frac{1}{\sigma_2/\sigma_1}     \right).
\label{ntot}
\end{equation}
Therefore, the values of the present-day surface densities $\sigma_2$ and  $\sigma_1$ to insert in eqs. (\ref{S2}) and (\ref{S1}), respectively are the following ones,
\begin{equation} 
 \sigma_2= \frac{ \sigma_{tot}} {\left( 1+\frac{1}{\sigma_2/\sigma_1}     \right)} \mbox{  and  } \, \sigma_1 =\sigma_{tot}- \sigma_2.
\label{newD}
\end{equation}

\begin{figure*}
\begin{centering}
\includegraphics[scale=0.37]{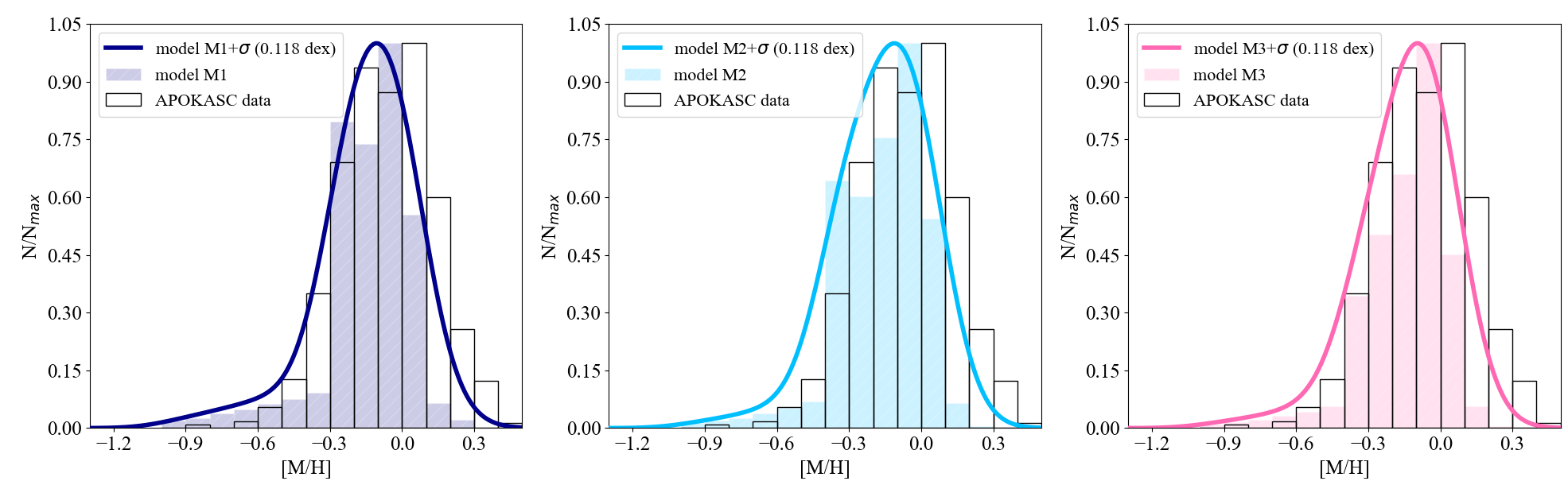}
\caption{Metallicity distributions predicted by models M1 (left panel), M2 (middle panel) and M3 (right panel) with the best fit model parameters (colored histograms). The observed APOKASC distribution calculated including both high-$\alpha$ and low-$\alpha$ stars is shown by the black, empty histograms. The solid  lines indicate the metallicity distribution of our chemical evolution models convolved with a Gaussian with standard deviation $\sigma$ = 0.118 dex (average APOKASC data error).  In each plot the  distributions are normalised to the corresponding maximum number of stars $N_{max}$.}
\label{MDF2}
\end{centering}
\end{figure*}

\begin{figure}
\begin{centering}
\includegraphics[scale=0.37]{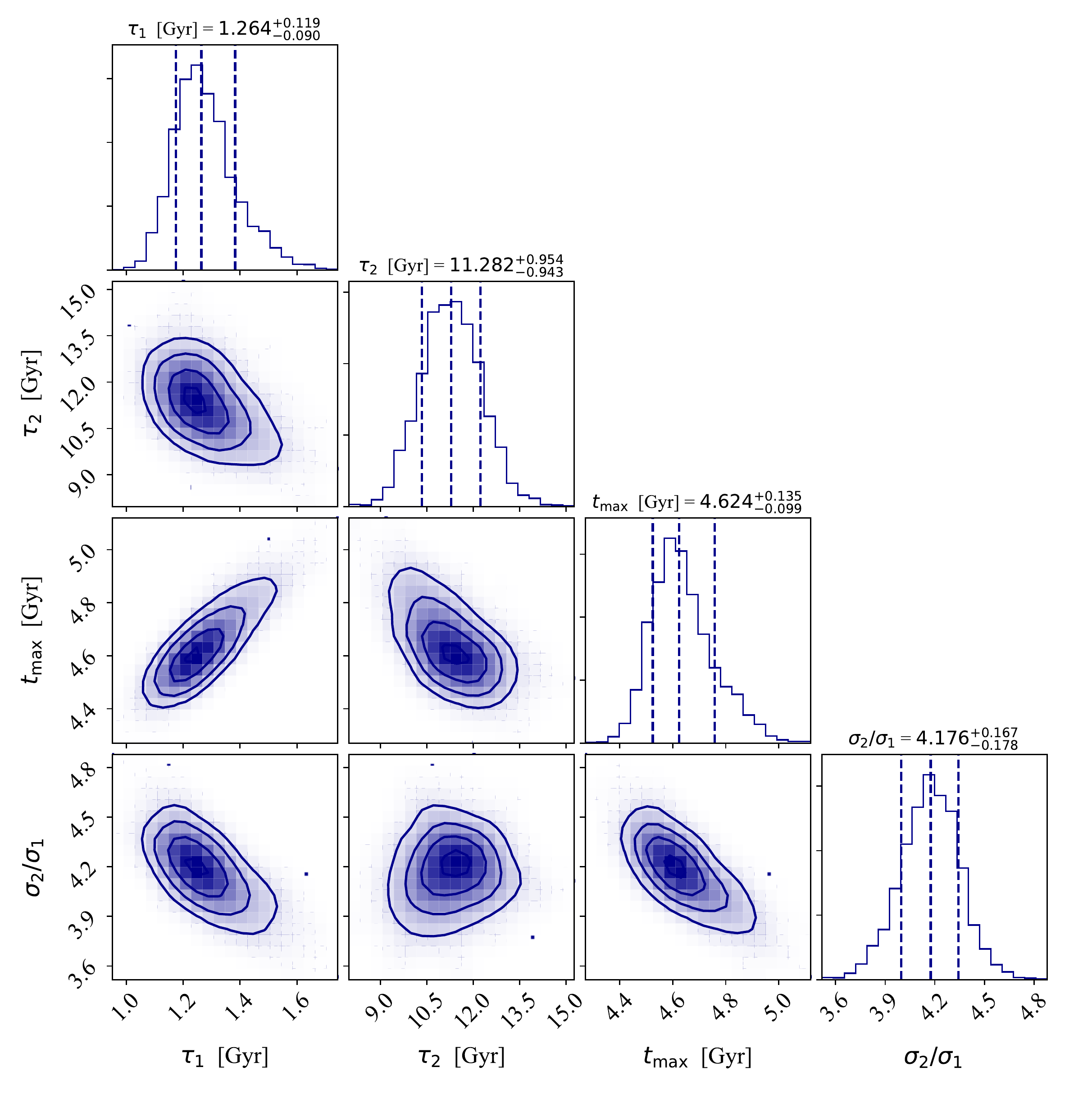}
\caption{Corner plot showing the posterior PDFs of  model M2 in which we adopt different SFEs for the high-$\alpha$ and low-$\alpha$ sequences: $\nu_1=2$ Gyr$^{-1}$ and $\nu_2=1$ Gyr$^{-1}$. The best fit of the four dimensional parameter space ${\bf \Theta} = \{\tau_1, \ \tau_2, \ t_{\rm max}, \ \sigma_2 / \sigma_1\}$  is  obtained by fitting [$\alpha$/Fe], [Fe/H] and ages
  of the APOKASC sample (see model details in Section \ref{SFE_var}). The median, 16$^{\rm th}$  and 84$^{\rm th}$ percentiles of
  the posterior PDF are plotted for each parameter above the marginalised PDF. }
\label{4par_corner_M2}
\end{centering}
\end{figure}

We consider, as the reference case, the model with the   SFE  fixed at  the value of $\nu= 1.3$ Gyr  (model M1)   as assumed in  ES19 where the predicted solar values for Mg, Fe, and Si were in agreement within one sigma with \citet{asplund2005} values (see their Table 1). 
In  Fig. \ref{4par_corner_M1} we show the corner plot with the posterior PDFs of the model M1 characterized by the  four model parameters  ${\bf \Theta} = \{\tau_1, \ \tau_2, \ t_{\rm max}, \ \sigma_2 / \sigma_1\}$ with priors as introduced in Section \ref{fitting}. 
 We still find  a significant delay in the start of the second gas infall, even larger than the the value found in ES19, $t_{{\rm max}}=5.278^{+0.261}_{-0.182}$ Gyr. The best model  predicts  for the $\sigma_2/\sigma_1$  ratio a value of  3.472$^{+0.234}_{-0.278}$. Therefore, our analysis 
favours the value derived by \citet{fuhr2017}, whereas the much larger value suggested by \citet{nesti2013} seems unsuitable to reproduce the APOKASC data. 
Moreover,  in the [$\alpha$/Fe] versus [Fe/H] space, this model presents results definitely in agreement with the finding of ES19.
Based on a  statistical method, we have full confirmation of a significant delay between the two infall episodes, as shown in Fig.
\ref{alpha_fe_4param_M1}. 

We can see  from Table \ref{tab1}   that model M1 predicts  Fe solar abundance within $3\sigma$,   Si   within $2\sigma$ and Mg within $1\sigma$  of the observational estimates by \citet{asplund2005}. The model presented in ES19 was able to reproduce solar values for the above mentioned elements within $1\sigma$. Now we predict  smaller  solar values compared to ES19 ones because  of the longer best fit  time-scales of accretion:   i.e. $\tau_1=1.112^{+0.215}_{-0.145}$ Gyr (first infall) and $\tau_2=13.596^{+1.874}_{-1.776}$ Gyr (second infall). Hence,  the chemical enrichment  for the model M1 is less efficient and evolves slower in time, leading to smaller solar values compared to ES19.

We also notice that the "loop" in the low-$\alpha$ sequence
does not cover all data. We remind the reader that we are considering a model  designed for the solar neighborhood
and we do not include stellar migration effects.
In principle, stellar migration \citep{schoenrich2009MNRAS} can help in reproducing the low-$\alpha$ sequence  composed by stars  with the smallest  [Fe/H] values (with stars migrating into the solar neighbourhood from the outer disk), as well as with the largest   [Fe/H] values (with stellar migration from inner  disc regions).

\begin{figure}
\begin{centering}
\includegraphics[scale=0.43]{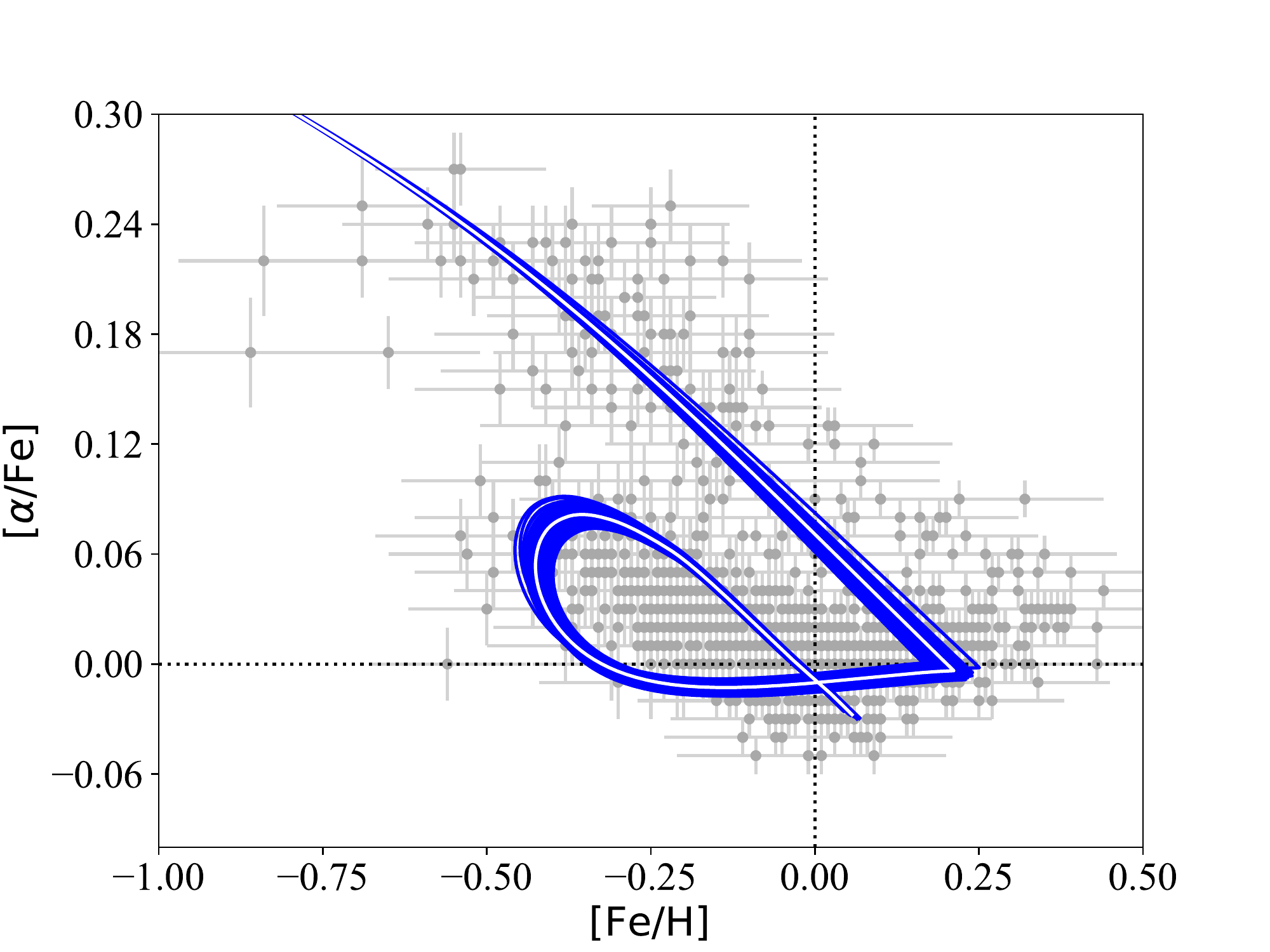}
\includegraphics[scale=0.43]{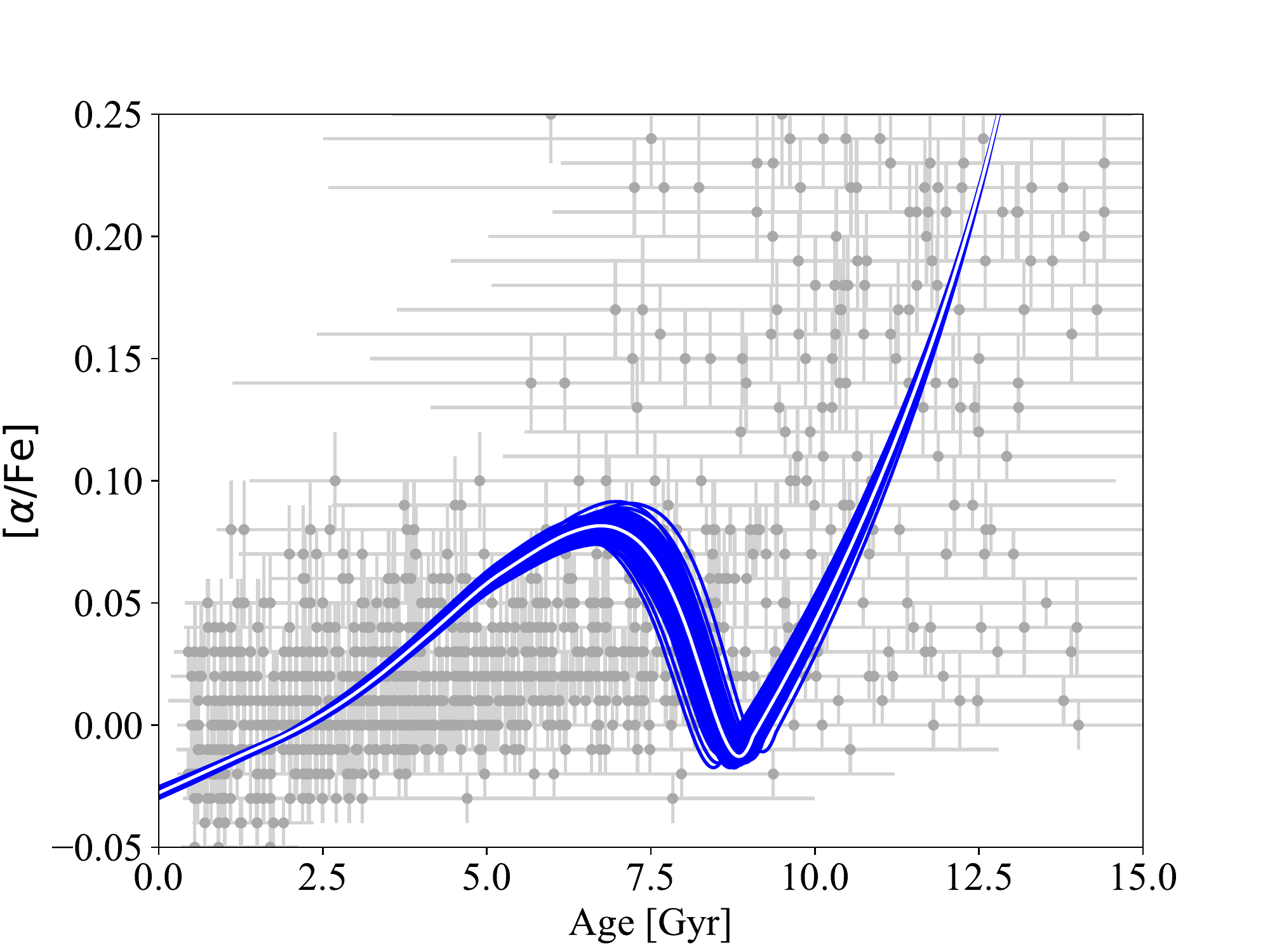}
\includegraphics[scale=0.43]{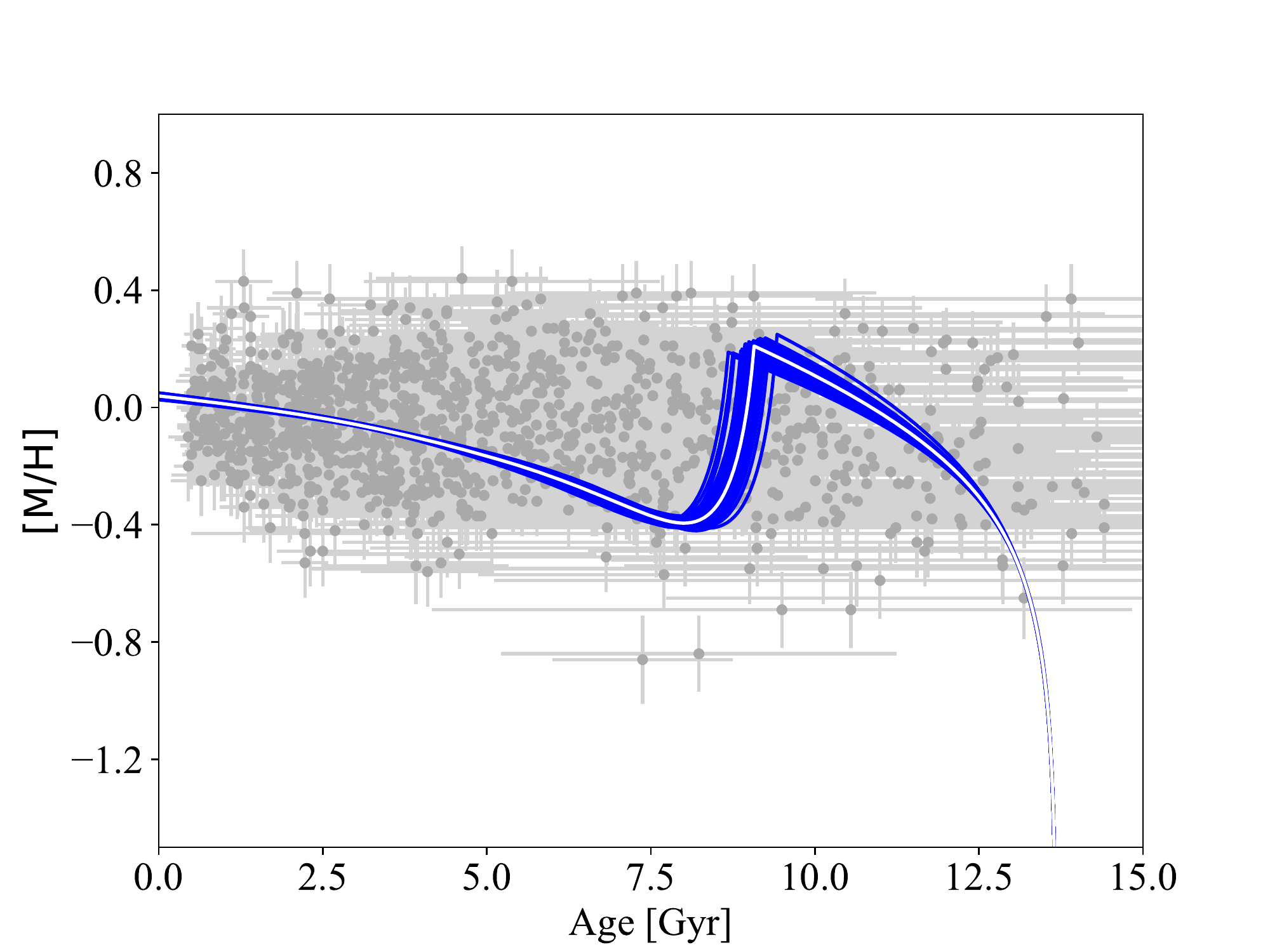}
\caption{Same as Fig. \ref{alpha_fe_4param_M1} for model M2 (see  in Section \ref{SFE_var}).} 
\label{alpha_fe_4param_M2}
\end{centering}
\end{figure}

\begin{figure}
\begin{centering}
\includegraphics[scale=0.37]{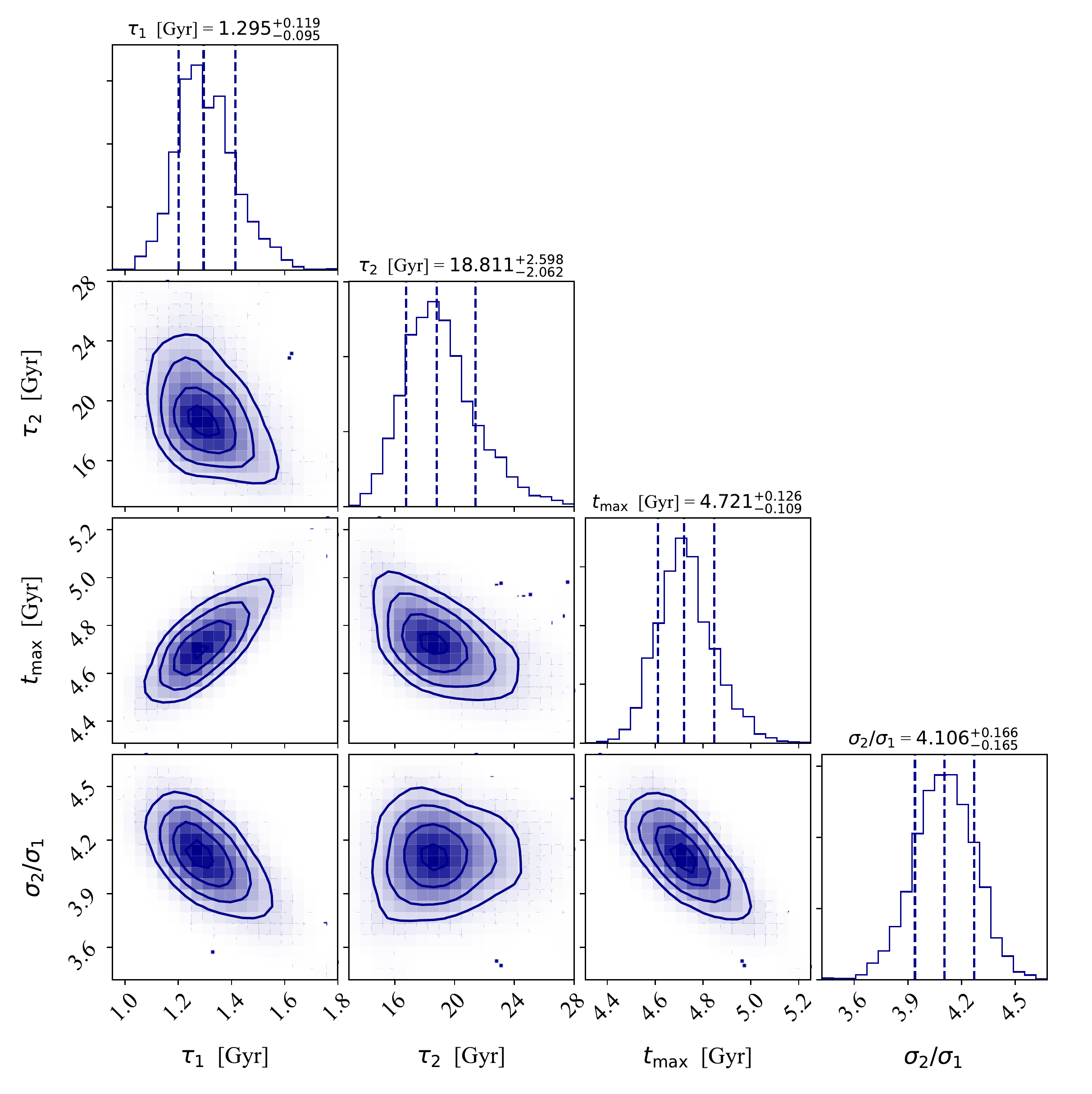}
\caption{Corner plot showing the posterior PDFs of  model M3 in which  we adopt different SFEs for the high-$\alpha$ and low-$\alpha$ sequences: $\nu_1=2$ Gyr$^{-1}$ and $\nu_2=1.3$ Gyr$^{-1}$ (see model details in Section \ref{SFE_var}). The median, 16$^{\rm th}$  and 84$^{\rm th}$ percentiles of
  the posterior PDF are plotted 
  for each parameter above the marginalised PDF. }
\label{4par_corner_M3}
\end{centering}
\end{figure}

\begin{figure}
\begin{centering}
\includegraphics[scale=0.43]{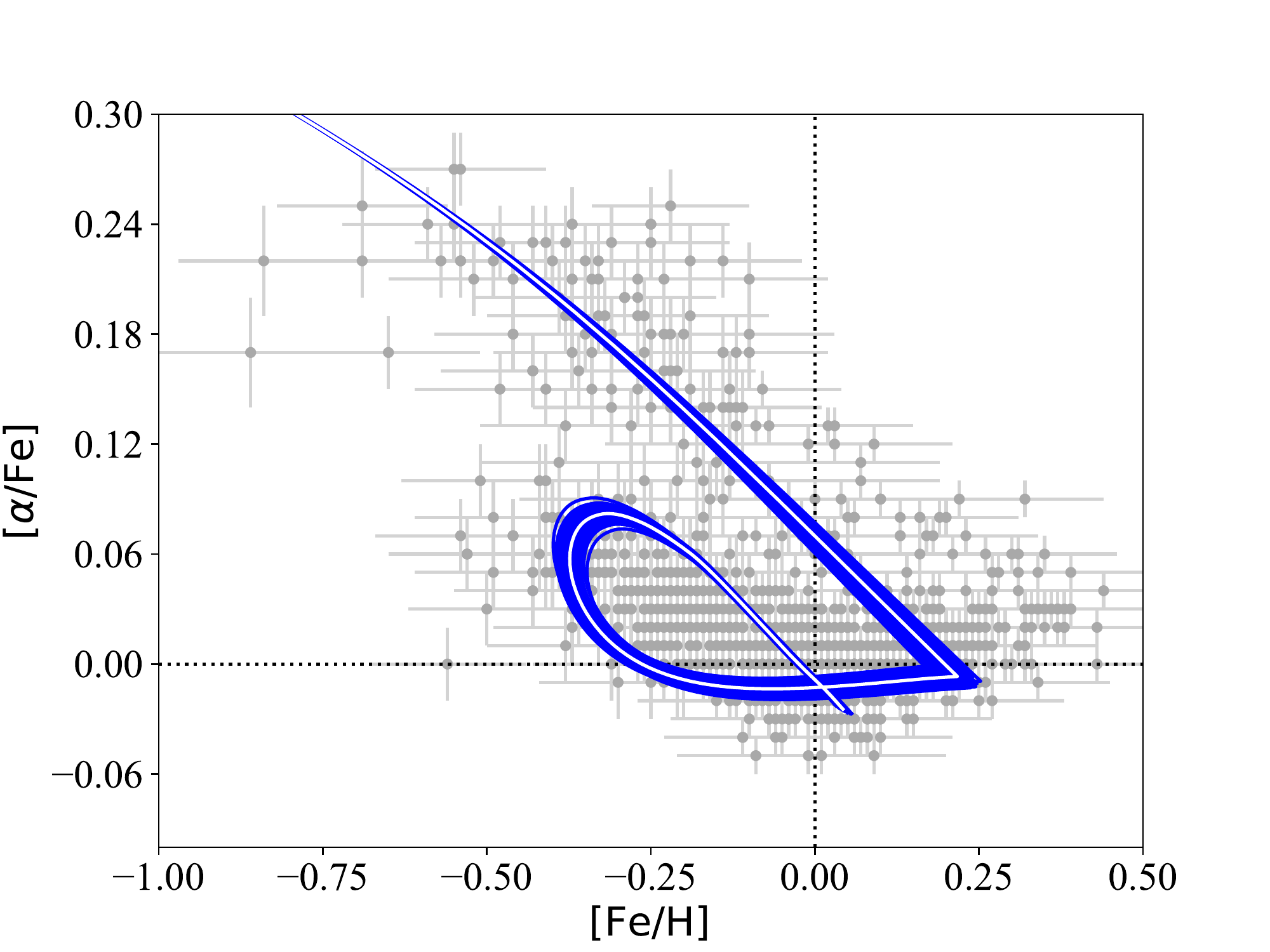}
\includegraphics[scale=0.43]{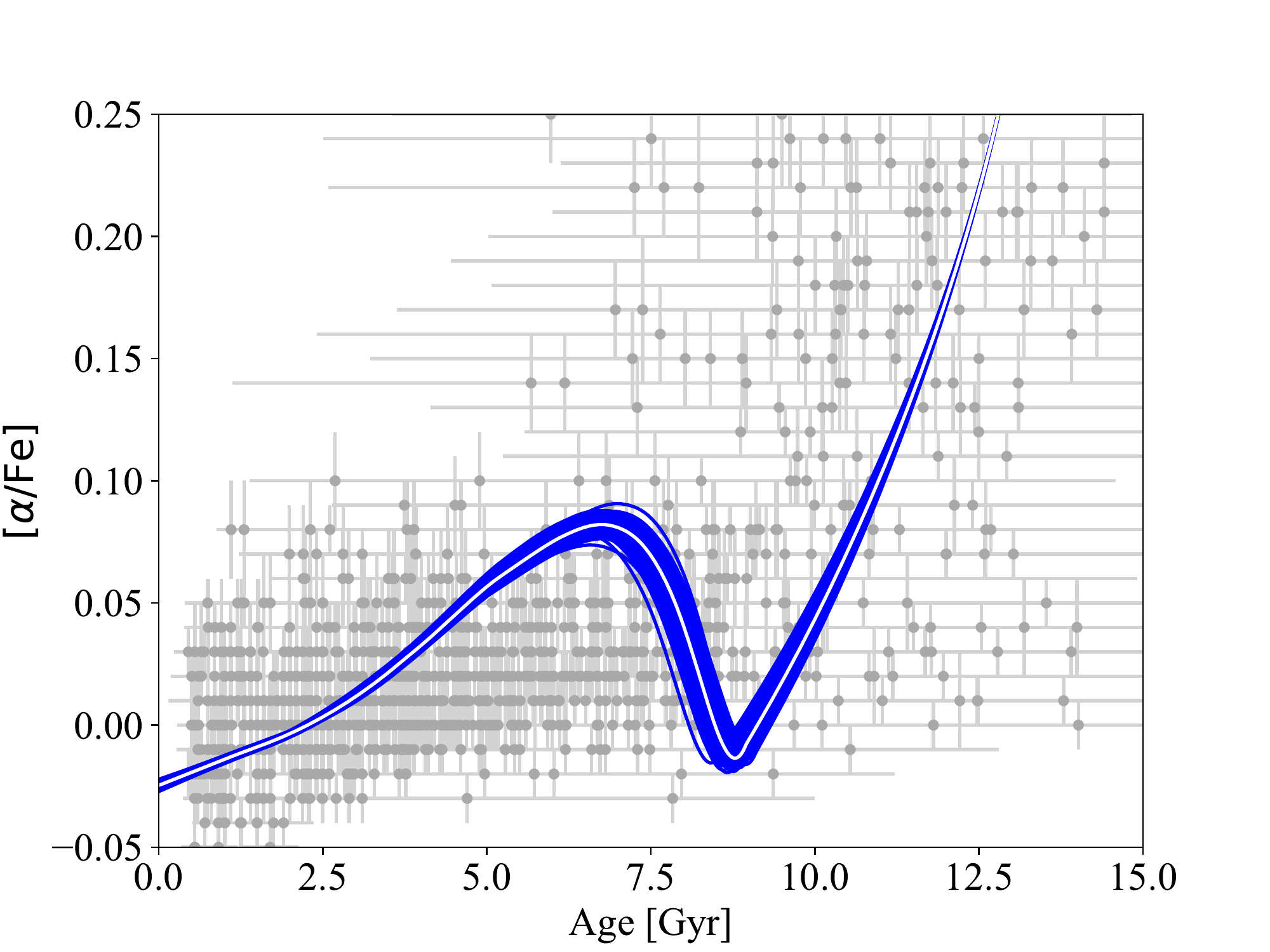}
\includegraphics[scale=0.43]{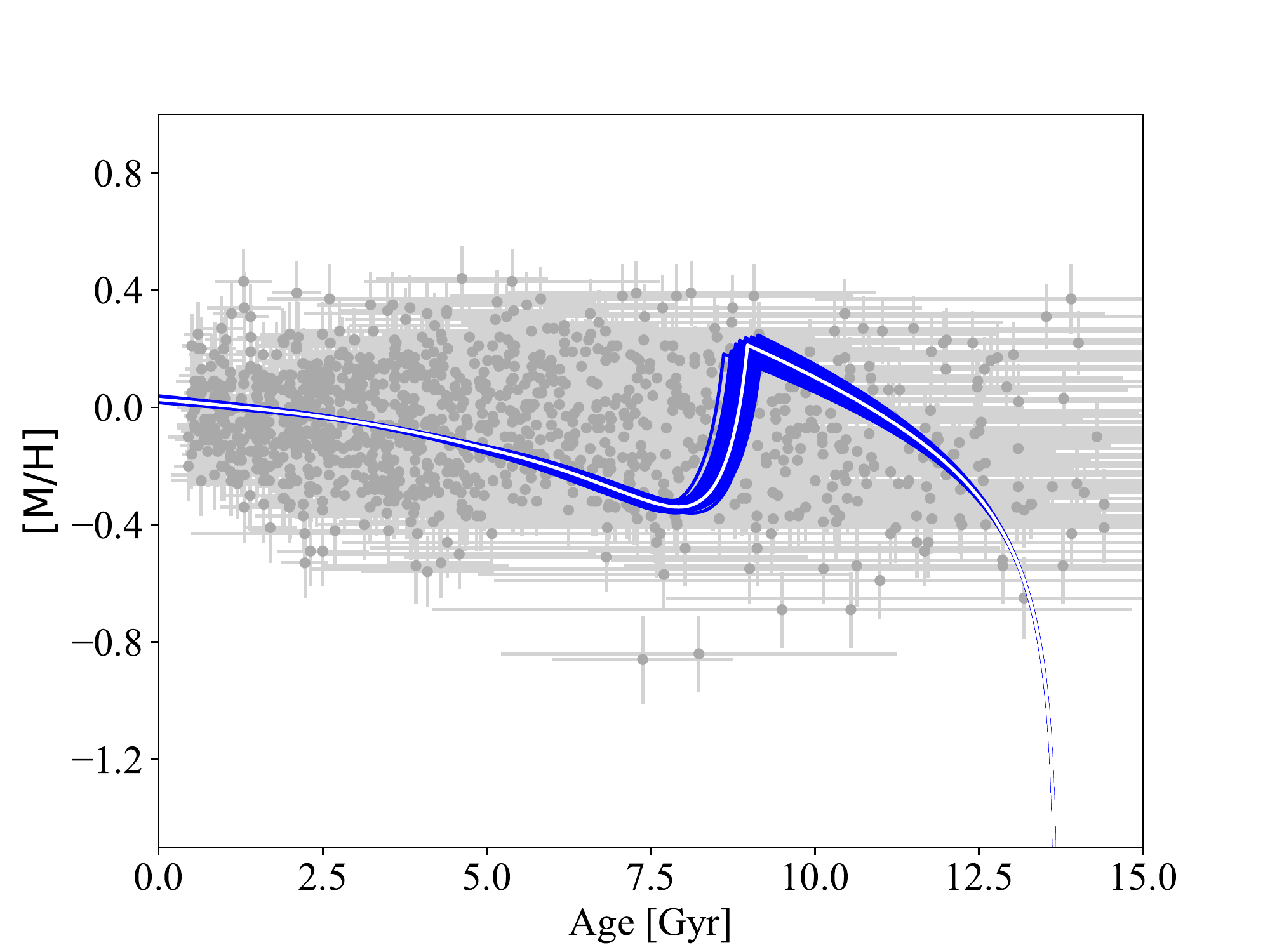}
\caption{Same as Fig. \ref{alpha_fe_4param_M1} for model M3 with  SFEs  fixed at the values of $\nu_1=2$ Gyr$^{.1}$  $\nu_2=1.3$ Gyr$^{.1}$  (see model details in Section \ref{SFE_var}).} 
\label{alpha_fe_4param_M3}
\end{centering}
\end{figure}

In Fig. \ref{alpha_fe_4param_M1},  the temporal  evolution of the  [$\alpha$/Fe] shows   an evident  "bump"   and the age-metallicity [M/H] \footnote{In the age-metallicity relation,   the metallicity [M/H] is computed using the expression introduced by \citet{salaris1993}, as done in ES19 to be consistent with the APOKASC sample:

\begin{equation} 
 \mbox{[M/H]}   = \mbox{[Fe/H]}  + \log \left( 0.638 \times
   10^{[\alpha/{\rm Fe}]}+ 0.362 \right).
\label{MH}
\end{equation}} relation shows a  sudden "drop",  both signatures of the delayed infall of gas in agreement with ES19 predictions.
The presence of such features is not obvious   in the observations but could be hidden behind the observational uncertainties.
Here we have tested    that    Bayesian methods lead to best models characterized by  significant delay and  important gas dilution. 

We note that [$\alpha$/Fe] values for the youngest stars tends to fall above the predicted [$\alpha$/Fe] vs. age trend. 
The predicted slope of the [$\alpha$/Fe] vs age relation for the low-$\alpha$  disc   is similar to the one presented by \citet{chiappini2015} and agrees in fact better with the trend  found by \citet{nissen2016} for solar twin  stars in the solar neighborhood than with the APOKASC data.

An important constraint for the chemical evolution model  is  the present-time stellar surface mass density.
The left panel of  Fig. \ref{SFR_SN}  shows that our best model predicts a value of 33.28  M$_{\odot} \mbox{ pc}^{-2}$, in excellent agreement with   the value of  33.4 $\pm$ 3 M$_{\odot} \mbox{ pc}^{-2}$ proposed by \citet{mckee2015}.
From middle panel of  Fig. \ref{SFR_SN} we notice that the  predicted present-day  SFR value  of 4.08 M$_{\odot}$ pc$^{-2}$ Gyr$^{-1}$ is in agreement with the measured range in the solar vicinity of 2-5 M$_{\odot}$ pc$^{-2}$ Gyr$^{-1}$  \citep{matteucci2012,prantzos2018}.

 \begin{table*}[htp]
\begin{center}
\tiny
\caption{Predicted delay $t_{{\rm max}}$, present-day surface density
  ratio $\sigma_{2}$/ $\sigma_{1}$, and infall time-scale $\tau_1$
  values by the  models  M1, M2 and M3  presented in this work (see text for model details). In the last column, we also provide the range values admitted by our study. The assumed values for high-$\alpha$  ($\nu_1$) and low-$\alpha$  ($\nu_2$)  SFEs for the different models are also indicated. }
\label{tab2}
\begin{tabular}{c|ccc|c}
\hline
  \hline
 &  &{\it Models} &&\\
 &  &&&\\

  & M1&M2&M3&\\
  
\hline
&&&\\
  $\nu_1$  [Gyr$^{-1}$]  &1.3&2.0&2.0  \\
  &&&\\
   $\nu_2$  [Gyr$^{-1}$]   &  1.3&1.0&1.3\\
   &&&\\
\hline

    & &{\it MCMC Results}&&Range\\
      & &&&\\
 $t_{{\rm max}}$ [Gyr]&5.278$^{+0.261}_{-0.182}$&4.624$^{+0.135}_{-0.099}$& 4.721$^{+0.126}_{-0.109}$&$4.525-5.539$\\

&&&\\

 $\sigma_{2}$/ $\sigma_{1}$&3.472$^{+0.234}_{-0.278}$&4.176$^{+0.167}_{-0.178}$&  4.106$^{+0.166}_{-0.165}$&$3.194-4.334$\\
 
 &&&\\

 $\tau_{1}$ [Gyr]&1.112$^{+0.215}_{-0.145}$&1.264$^{+0.119}_{-0.090}$&  1.295$^{+0.119}_{-0.095}$& $0.967-1.414$\\
 
  &&&\\
  
  $\tau_{2}$ [Gyr]&13.596$^{+1.874}_{-1.776}$&11.282$^{+0.954}_{-0.943}$&  18.811$^{+2.598}_{-2.062}$& $10.339-21.409$\\
 
  &&&\\
 \hline
\end{tabular}
\end{center}
\end{table*}

The time evolution of the Type Ia SN and Type II SN rates are also plotted   in  Fig. \ref{SFR_SN}. The present-day Type II SN rate in the whole Galactic disc predicted by our model is 1.67 /[100 yr], in good agreement with the observations of \citet{li2010} which yield a value of 1.54 $\pm$0.32 /[100 yr]. 
The predicted present-day Type Ia SN rate in the whole Galactic disc
is 0.34 /[100 yr], again in good agreement with the value provided by \citet{cappellaro1997} of 0.30$\pm$0.20  /[100 yr].

In the left panel of  Fig. \ref{MDF2} we compare the metallicity distribution function (MDF)  of the model M1  with the whole APOKASC data sample.  Although  the  predicted MDF is consistent with the data,  it underestimates  the number of stars at super-solar metallicities.  This is due to the longer best fit time-scales of accretion compared to the classical "two-infall" model.
In  Fig. \ref{MDF2} we also draw the curve related to the model distribution convolved with a Gaussian with a  constant dispersion  fixed at the value of $\sigma= 0.118$ dex, which is the average [M/H]  observational  error in APOKASC data (see ES19). 
In this  case we improve the fit and the high metallicity tail of the MDF is  better accounted for.

In the next Section we will test how the delay between the two infall episodes is sensitive to the choice of the SFE parameter.

\subsection{SFE parameter study (models M2 and M3)} \label{SFE_var}
In this Section  we consider  different  SFE values as already used in previous works.
Different infall episodes could in principle  be characterized by different SFEs as suggested by  \citet{grisoni2017,grisoni2018}. In their chemical evolution models,  
the SFEs of the high-$\alpha$ and low-$\alpha$ sequences have been fixed at the values of  $\nu_1 = 2$ Gyr$^{-1}$ and  $\nu_2 = 1$ Gyr$^{-1}$, respectively.

 In  model M2 we adopt  the same prescriptions as \citet{grisoni2017,grisoni2018}  as shown in Table  \ref{tab2}, whereas in model M3 we consider $\nu_1 = 2$ Gyr$^{-1}$ and    $\nu_2 = 1.3$ Gyr$^{-1}$ (high-$\alpha$ SFE as \citealt{grisoni2017} and low-$\alpha$ one as ES19).

 In Fig. \ref{4par_corner_M2} the corner plot  of the posterior PDFs of model M2 confirms the trend  mentioned above with model M1. The best value for the  time delay  is $t_{{\rm max}}=4.624^{+0.135}_{-0.099}$ Gyr,
 and  $\sigma_2/\sigma_1=4.176^{+0.167}_{-0.178}$. 
This shows that  the  time-scales of accretion $\tau_1$ and $\tau_2$ are sensitive to the assumed SFE. During the  high-$\alpha$ phase, the best model  M2 is characterized by  a longer time-scale $\tau_1$  than the M1 one. 
In order to obtain a  chemical enrichment history similar to the one of the M1 model, an increase of the SFE must be  compensated by a longer  time-scale $\tau_1$; the same applies to the second infall timescale. 

 \begin{figure*}
\begin{centering}
\includegraphics[scale=0.5]{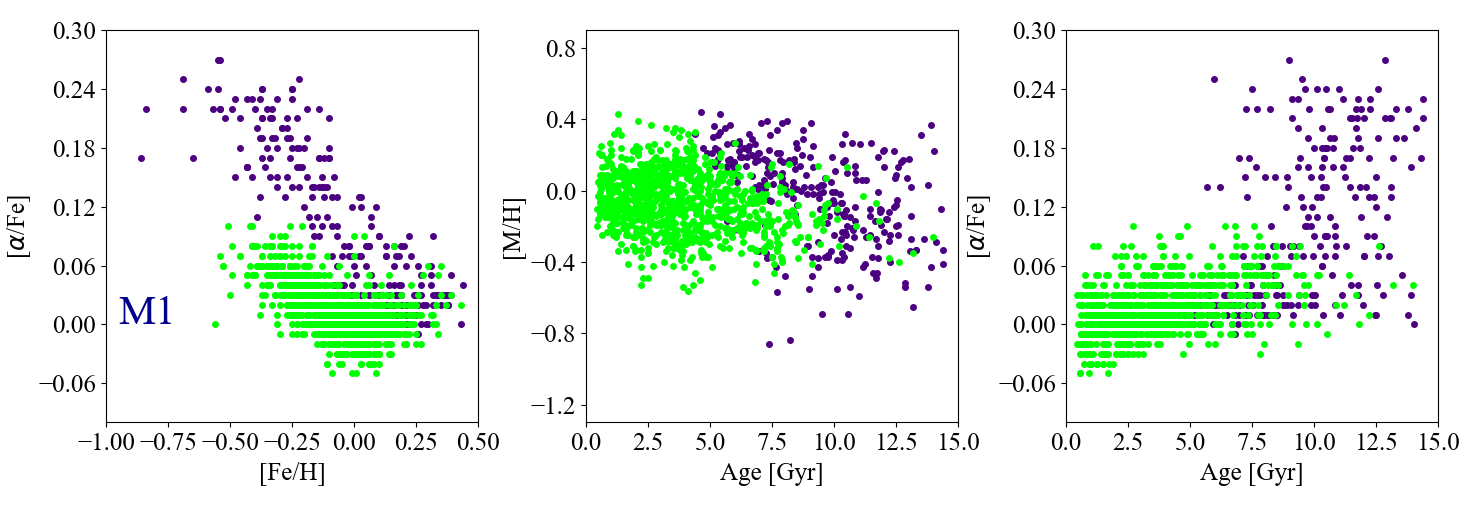}
\caption{Disc dissection of the APOKASC sample of \citet{victor2018} in high-$\alpha$ (violet points) and low-$\alpha$ (green points) sequence stars   based on the chemical evolution model M1.
In the left panel the abundance ratios  [$\alpha$/Fe] versus [Fe/H] for the APOKASC sample is reported. The age metallicity relation and the temporal evolution of the [$\alpha$/Fe] are shown in the middle and right panels, respectively. }
\label{dissection}
\end{centering}
\end{figure*}  
 In Fig.  \ref{alpha_fe_4param_M2}, it
  is possible also in  this case to appreciate the dilution effect of the gas rich accretion events in the [$\alpha$/Fe] versus [Fe/H] abundance ratios, in the age metallicity relation and in  [$\alpha$/Fe] versus age plot. By comparing Fig.  \ref{alpha_fe_4param_M2} with Fig.  \ref{alpha_fe_4param_M1} we note that the two models  show similar results.

 An important constraint for the chemical evolution model  is represented by the present-time stellar surface mass density.
The left panel of  Fig. \ref{SFR_SN}  shows that our best model  M2 predicts  a present-time stellar surface mass density  value of 32.60  M$_{\odot} \mbox{ pc}^{-2}$, which is slightly smaller than   the value proposed by \citet{mckee2015}. 
The   predicted present-day  SFR value  of 3.72 M$_{\odot}$ pc$^{-2}$ Gyr$^{-1}$ is smaller  than the one predicted by M1 model,  in better agreement with  the  observed range of 2-5 M$_{\odot}$ pc$^{-2}$ Gyr$^{-1}$  \citep{matteucci2012,prantzos2018}.

The present-day  Type Ia  and II SN rates  are also shown  in  Fig. \ref{SFR_SN}. The present-day Type II SN rate in the whole Galactic disc predicted by our model is 1.53 /[100 yr],  in good agreement with the observations by \citet{li2010}.
The predicted present-day Type Ia SN rate in the whole Galactic disc
is 0.33 /[100 yr],  in good  agreement with the value provided by \citet{cappellaro1997}.

In the middle panel of Fig. \ref{MDF2} we show the MDF for the model M2. We note that the convolution of the MDF  with a  gaussian of standard deviation equal to the typical [M/H] error in the APOKASC sample helps in reproducing the high- and low-metallicity tails of the observed distribution. Because of the larger SFE  value for the high-$\alpha$ sequence $\nu_1$, model M2 presents an MDF with more metal poor stars compared to the model M1.

Finally, we present the results  related to the model M3 characterized by  the following SFEs for the two gas infall episodes:   $\nu_1=2$ Gyr$^{-1}$ and
$\nu_2=1.3$ Gyr$^{-1}$.
The corner plot related to the model M3  can be found in Fig. \ref{4par_corner_M3}. The main difference between model M2 and model M3 is the time-scale of accretion of the second infall $\tau_2$; in fact the best fit model M3 requires  $\tau_2=18.811^{+0.126}_{-0.109}$ Gyr. However, as can be inferred from Figs. \ref{SFR_SN}, \ref{MDF2} and \ref{alpha_fe_4param_M3} no substantial differences characterize the chemical enrichment of the model M3 compared to models M1 and M2. The  model M3 shows a slightly larger solar abundance values (see Tabel \ref{tab1}), because of the higher SFEs in both high-$\alpha$ and low-$\alpha$ sequences.

In Table \ref{tab2} we summarize   the predicted delay $t_{{\rm max}}$, present-day surface density ratio $\sigma_{2}$/$\sigma_{1}$, infall time-scales $\tau_1$ and $\tau_2$ values obtained for  the different best fit models presented in this work.
It is clear that in all our tests, independently of  the SFE prescription,  the  presence of a  delay is a solid result. In fact, its  value spans the range $4.5-5.5$ Gyr.
Another important result is that we are capable to constrain the ratio $\sigma_{2}$/$\sigma_{1}$. The predicted values are in the range $3.2-4.3$,  in agreement with \citet{fuhr2017} and \citet{mac2017}.

 The predicted values for $\tau_1$, $t_{{\rm max}}$   and $\sigma_{2}$/$\sigma_{1}$ are not much sensitive to  different SFE prescriptions, confirming the robustness of the results. However,  the  best-fit accretion time-scale $\tau_2$ 
spans  a large range of values (10.3-21.4 Gyr) considering   models M1, M2 and M3. Hence, we cannot draw any firm conclusion  about this parameter  assuming only the  observational  constraints given by  abundance ratios and ages of the   APOKASC sample stars.

The predicted time-scales for the low-$\alpha$ sequence are substantially longer than the one proposed by the classical "two-infall" model by \citet{chiappini2001} and \citet{grisoni2018} for the solar neighborhood. In these works the "inside-out" formation scenario 
was obtained with an infall time-scale for the thin disc  that increases with the Galactocentric distance, and in particular in the solar neighborhood $\tau_2=7$ Gyr (3.3 Gyr smaller than  our lower limit predictions).
However,  the best fit  "low-$\alpha$ "time-scales of accretions for models M1 and M2 are in agreement with the    chemical evolution model proposed by  \citet{Nidever:2014fj}. In fact,   originally designed  to reproduce the APOGEE data, this  model is characterized by   an  $e$-folding time-scale of gas accretion  fixed at the value of 14 Gyr. 

  In a future work, it is our intention to  extend our results to other Galactocentric distances, analyzing the inside-out Galactic disc growth, with the inclusion  of  other observational constraints in the MCMC procedure.

\subsection{The dissection of the Galactic disc components}

In previous works, the Galactic disc dissection in the solar neighbourhood  was based either on the chemical tagging or using the  kinematics proprieties of the stars (see \citealt{victor2018} and references therein). 
In this Section we propose a new method to separate the APOKASC data in high-$\alpha$ and low-$\alpha$ disc components using the results of our best fit models M1, M2 and M3.
We present a new criterium in  which,  beside the chemical abundance of the stars, we also use  their asteroseimic age information.

Given a best fit model, we associate to each star 
in the  space of  observables  ${\bf x} = \{[\alpha/{\rm Fe}], \ [{\rm Fe}/{\rm H}], \ {\rm age}\}$,  
the closest point on the model hyper-surface using   eq.~(\ref{eq:distance2}) introduced in Section \ref{fitting}. 
If this point  on the  hyper-surface   is characterised by an age larger than delay $t_{\rm max}$, then the star is considered as part of the high-$\alpha$ sequence. On the other hand, if this age is smaller than the delay, then it is considered as part of the low-$\alpha$ sequence.

We found that all the models produce roughly the same disc separation, therefore  in Fig. \ref{dissection} we show only the results obtained with the M1 model. It is interesting to note that this  dissection criterium  produces a disc separation not much different from the one presented by \citet{victor2018}  based on chemistry (see their Fig. 8).
However, some relevant differences can be noted in the temporal evolution of the abundance ratio [$\alpha$/Fe] for old stars with  [$\alpha$/Fe$]< 0.05$ dex. The dissection based only on chemistry by \citet{victor2018}  tags all the stars in this region as low-$\alpha$ sequence. On the other hand, our separation--which uses the age information as well--predicts a mixed population of high-$\alpha$ and low-$\alpha$ stars close to the delay, $t_{\rm max}$.

By means of a chemo-dynamical model for the Milky Way, it will be possible to include also the kinematic information along with the stellar ages and chemical abundances in the  Bayesian analysis based on MCMC methods to better constrain the disc dissection and shed more light to the different disc components.

\section{Conclusions}\label{conc}

For the first time, we used  a detailed Bayesian analysis  to constrain chemical evolution models with stellar abundances and precise stellar ages provided by asteroseismology of the APOKASC sample by \citet{victor2018}.
We tested the robustness of the findings of \citet{spitoni2019} concerning  the importance of a
significant delay between the first infall and the start of the second one in the framework of the two-infall chemical evolution model, in order to reproduce the APOKASC sample in the solar annulus.  

In our analysis we considered four free parameters  (accretion  time-scales $\tau_1$ and $\tau_2$, delay $t_{\rm max}$ and present-day surface mass density ratio  $\sigma_2 / \sigma_1$). We tested three different SFE recipies: in model M1, SFE is fixed at the value of 1.3 Gyr$^{-1}$ during the Galactic life, following \citet{grisoni2017} in model M2 and M3  the high-$\alpha$  and low-$\alpha$ sequences are characterized by different SFEs.

Our main conclusions can be summarized as follows:
\begin{itemize}

\item  The best fit models M1, M2 and M3 present  a delay  between the two infall episodes in agreement with \citet{spitoni2019}.
 These models  also reproduce reasonably 
well other important observational constraints for the chemical evolution of the disk, including the present-day  stellar surface  density by \citet{mckee2015},  Type II and Type Ia SN rates, the  SFR, the metallicity distribution function of the APOKASC data and the solar abundance values of \citet{asplund2005}.

\item We have shown with a Bayesian analysis that the  presence of a consistent delay is robust against the uncertainties in the SFEs, and the value lies in the range $4.5-5.5$ Gyr for different models. 

\item The best fit model parameter for  present-day surface mass density ratio $\sigma_2/\sigma_1$  between low-$\alpha$ and high-$\alpha$ sequences
spans the range $3.2-4.3$, which is in agreement with the findings of \citet{fuhr2017}.

\item We used our best models to dissect the Galactic disc components of the APOKASC sample. The results of the dissection are 
similar to those presented by \citet{victor2018} based only on chemistry. 
 Differences in the disc separations are for the stars  close to the  model transaction between  high-$\alpha$ and low-$\alpha$ sequences  in the [$\alpha$/FE] versus [Fe/H] space.

\end{itemize}

Different physical reasons can be associated to a  significant delay in the range $4.5-5.5$ Gyr  between the two accretion   episodes.

In the two-infall model scenario coupled with  the shock-heating theory, a significant delay between the accretion phases has been suggested also by  \citet{noguchi2018} .  In their picture, 
a first infall episode originates the high-$\alpha$ sequence, which is followed by a hiatus  until the shock-heated gas in the Galactic dark matter halo has radiatively cooled and can be accreted by the Galaxy. 
In this framework, \citet{noguchi2018}  found that the  SFR of the Galactic disc is characterised by two peaks separated by  $\sim 5$ Gyr (in agreement with ES19 and our findings).

 The significant delay in the  two-infall model of ES19 
has been also discussed by \citet{vincenzo2019} in the context  of the stellar system accreted by the Galactic halo, AKA  Gaia-Enceladus  \citep{helmi2018,koppelman2019}. \citet{vincenzo2019} presented the first chemical 
evolution model for Enceladus, investigating the star formation history of one of the most massive satellites accreted 
by the Milky Way during a major merger event. It was proposed that the mechanism which quenched the Milky Way star formation at high redshift by heating 
up the gas in the dark matter halo was a major merger event with a satellite like Enceladus.  This proposed scenario is in agreement with the recent  \citet{chaplin2020} study. They constrained the merging time with the very bright, naked-eye star $\nu$ Indi finding that, at 68\% confidence,  the earliest the merger could have started  was 11.6 Gyr ago.

 Finally, the delay could be  interpreted as  the main effect  of  a  late gas-rich accretion episode which  shaped the low-$\alpha$ sequence, 
as confirmed in early works of chemical evolution in a cosmological context \citep{calura2009} and more recently by cosmological simulations \citep{buck2020}.

We are aware that our study is  limited to the solar annulus region, and that other dynamical processes such as  stellar migration \citep{schoenrich2009MNRAS}  might have played an important role during the Galactic evolution.

\section*{Acknowledgement}
 The authors thank the anonymous referee for various suggestions that improved the paper.
Funding for the Stellar Astrophysics Centre is provided by The Danish National Research Foundation (Grant agreement no.: DNRF106).
E. Spitoni thanks P. E. Nissen, A. Saro and  M. Fredslund Andersen for useful discussions.
E. Spitoni and V. Silva Aguirre acknowledge support from the Independent Research Fund Denmark (Research grant 7027-00096B). 

\bibliographystyle{aa} 
\bibliography{disk}

\end{document}